\documentclass[11pt]{article}
\usepackage{natbib}
\usepackage{graphics}
\usepackage{graphicx}
\usepackage{setspace} 
\usepackage[portrait, margin=1.25in]{geometry}
\usepackage{amsmath}  
\usepackage{amsfonts}
\usepackage{indentfirst}
\usepackage{array}
\usepackage{booktabs}
\usepackage{multirow}
\usepackage{booktabs}
\usepackage{tabularx}
\usepackage{tabulary}
\usepackage{dcolumn}
\usepackage{tabularht}
\usepackage{verbatim}
\usepackage{graphicx}
\usepackage{booktabs,caption}
\usepackage[flushleft]{threeparttable}
\usepackage{float}

\usepackage{enumitem}   
\usepackage{placeins}
\usepackage{authblk}
\usepackage{subcaption}

\usepackage{microtype}
\usepackage{amsmath,amssymb,amsthm,enumitem}
\newtheorem{assumption}{Assumption}
\newtheorem{proposition}{Proposition}[section]
\newtheorem{condition}{Condition}

\newtheorem{remark}{Remark}
\newcommand{\R}{\mathbb{R}}
\newcommand{\E}{\mathbb{E}}
\newcommand{\Pp}{\mathbb{P}}
\newcommand{\op}{\mathrm{op}}

\newcommand{\var}{\operatorname{Var}}

\newcommand{\iid}{\mathrm{i.i.d.}}

\usepackage{xcolor}

\definecolor{dark-maroon}{HTML}{5D0F0D}
\definecolor{navyblue}{HTML}{0A2044}

\definecolor{purple}{HTML}{5601A4}
\definecolor{navy}{HTML}{0D3D56}
\definecolor{ruby}{HTML}{9a2515}
\definecolor{alice}{HTML}{107895}
\definecolor{daisy}{HTML}{EBC944}
\definecolor{coral}{HTML}{F26D21}
\definecolor{kelly}{HTML}{829356}
\definecolor{cranberry}{HTML}{E64173}
\definecolor{jet}{HTML}{131516}
\definecolor{asher}{HTML}{555F61}
\definecolor{slate}{HTML}{314F4F}

\usepackage{hyperref}
\hypersetup{
    colorlinks= true,
    citecolor= dark-maroon,
    linkcolor= dark-maroon,
    filecolor= dark-maroon,      
    urlcolor= dark-maroon,
}

\DeclareMathOperator*{\argmin}{arg\,min}

\makeatletter

\makeatletter
\tagsleft@false
\@ifundefined{c@journalappendixequation}{\newcounter{journalappendixequation}}{}
\@ifundefined{journalappendixmath}{%
  \newenvironment{journalappendixmath}{%
    \begingroup
    \edef\journalmainequation{\number\value{equation}}%
    \setcounter{equation}{\value{journalappendixequation}}%
    \ifdefined\theHequation
      \renewcommand{\theHequation}{appendix.A.\arabic{equation}}%
    \fi
  }{%
    \setcounter{journalappendixequation}{\value{equation}}%
    \setcounter{equation}{\journalmainequation}%
    \endgroup
  }%
}{}
\makeatother
\renewcommand{\qedsymbol}{\ensuremath{\blacksquare}}
\providecommand{\norm}[1]{\left\lVert#1\right\rVert}

\makeatletter
\tagsleft@false
\@ifundefined{c@journalappendixequation}{\newcounter{journalappendixequation}}{}
\@ifundefined{journalappendixmath}{%
  \newenvironment{journalappendixmath}{%
    \begingroup
    \edef\journalmainequation{\number\value{equation}}%
    \setcounter{equation}{\value{journalappendixequation}}%
    \ifdefined\theHequation
      \renewcommand{\theHequation}{appendix.A.\arabic{equation}}%
    \fi
  }{%
    \setcounter{journalappendixequation}{\value{equation}}%
    \setcounter{equation}{\journalmainequation}%
    \endgroup
  }%
}{}
\makeatother
\renewcommand{\qedsymbol}{\ensuremath{\blacksquare}}

\begin{document}

\onehalfspacing

\title{Uniformly Valid Inference Under Interactive and High-Dimensional Constraints }
\author{Joseph Fry}  
\affil{\footnotesize
Rutgers University, 75 Hamilton St, 08901, New Brunswick, New Jersey, United States\par\nopagebreak
 \textit{E-mail address:} \texttt{jf1372@economics.rutgers.edu}}
\date{\today}
\maketitle

\begin{abstract}
    Asymptotic normality approximations often fail to hold for extremum estimators when the true value of the parameter is at or close to the boundary of a parameter space. I analyze and develop tests using a quasi-unconstrained estimator, which is asymptotically normal even when the true parameter vector is near or at the boundary. These results generalize previous work with this estimator by allowing interactions between when different parameters are close to the boundary and showing how the method can naturally be modified when a nuisance parameter is also high-dimensional. I show that variations of Wald, Likelihood Ratio, and Lagrange Multiplier tests can control size in a uniform sense, provided the initial constrained estimator is sufficiently accurate. I verify the high-level conditions in a correlated random coefficient model and a model with monotonic treatment effects.
\end{abstract}

\section{Introduction} \label{s: Intro}

General asymptotic normality results for unconstrained extremum estimators are well-established in the literature (see, e.g., \cite{Newey-McFadden}). Similarly, when the true value of the parameter is fixed at some interior point of a constrained parameter space, then the constraints are non-binding with probability approaching one and the constrained estimator is asymptotically equivalent to the unconstrained estimator. However, \cite{Andrews1999} and \cite{ConstrainedMEstimation} show that when the parameter is fixed at a boundary point of the parameter space, the asymptotic distribution of the constrained estimator is equal to the distribution of a normal random vector that has been projected onto the tangent cone of the feasible set at that boundary point. The fact that the asymptotic distribution is discontinuous in the parameter results in it providing poor approximations in finite samples when the parameter is close to the boundary of the parameter space relative to the sample size, as can be illustrated by examining the asymptotics with sequences of the parameter drifting towards the boundary. Here, I generalize the results of \cite{KETZ2018}, which builds on earlier work using quadratic approximations by \cite{Andrews1999}, to show that conducting inference using a quasi-unconstrained estimator can allow us to control size in a uniform sense, meaning that the procedure can control size in large samples regardless of how close to the boundary the parameter is.
\par 
\cite{KETZ2018} uses what they refer to as a quasi-unconstrained estimator (QUE) to calculate test statistics. This is calculated by taking some $\sqrt{n}$-consistent constrained estimator and performing a single Newton–Raphson iteration. If the true value of the parameter is a local minimum on the unconstrained parameter space, then this quasi-unconstrained estimator will also be $\sqrt{n}$-consistent. Furthermore, since it satisfies a set of first-order conditions asymptotically, it is asymptotically normal by similar reasoning as for an unconstrained extremum estimator. \cite{KETZ2018} then uses the QUE to conduct inference on subvectors using a Conditional Likelihood Ratio (CLR) test based on the conditioning principle of \cite{Moreira2003}. They show that this test will control asymptotic size in a uniform sense, but only when the parameter space meets specific conditions. They assume the parameter space is a product space, so each element of the parameter vector is constrained to lie in some closed interval (e.g., $\theta_j \in [0,c]$ for some constant $c$ or $\theta_j \in [0, \infty)$). While this covers important applications, such as estimating parameters subject to sign restrictions, it excludes cases where constraints interact. For example, autoregressive models restricted to be stationary, estimation of consumer demand systems in which symmetry and negative semidefiniteness are imposed on the Slutsky matrix, and a random coefficient model with correlated coefficients. This also arises when parameters are constrained to be convex combinations, such as in Markov transition models, mixture models and the synthetic control method. Allowing for the parameter space to have a non-product structure, I use the results of \cite{ANDREWS2020} to show that the Wald and Conditional Lagrange Multiplier (CLM) subvector tests are uniformly asymptotically equivalent and satisfy uniform asymptotic similarity. For the CLR test, I show that, with a slight modification, it can control size uniformly but will not be similar in general. This potential for the CLR to be conservative has implications for power, and I provide examples of how the Wald test may dominate the CLR or vice versa. The uniformity results are established under high-level sufficient conditions for the asymptotic normality of the QUE under drifting sequences. I verify these high-level conditions in a correlated random coefficient model and a model where treatment effects are constrained to be monotonic.
\par
A substantial literature has developed on inference when the parameter lies on or near the boundary of the parameter space. \cite{TestingOnTheBoundary} provides methods for when the parameters fixed by the null hypothesis may lie on the boundary of the parameter space and there may be a nuisance parameter that appears under the alternative hypothesis, but not under the null. \cite{LRTestOnTheBoundary} give the asymptotic distribution of the Likelihood Ratio test when the parameter is fixed on the boundary of the parameter space. However, in practice, it is generally unknown whether the parameter is on the boundary or close to it, and they do not address the case where the parameter is close to the boundary relative to the sample size. Standard resampling methods for inference also suffer from limitations when the parameter is close to or at the boundary. \cite{BootstrapInconsistency} shows that the standard bootstrap can be inconsistent when the parameter is on the boundary of the parameter space, and \cite{ProblemWithSubampling} show that subsampling and $m$ out of $n$ bootstrap methods do not control asymptotic size in a uniform sense when the parameter may be close to the boundary. Intuitively, this is because the shape of the finite-sample distribution depends on how close the parameter is to the boundary relative to the sample size, which differs when using the full sample versus subsamples. It is possible to modify existing tests to ensure asymptotic size control and/or to increase local asymptotic power by using adaptive critical values, but these modifications typically are computationally expensive (e.g., \cite{MCCLOSKEY2017}) or specialized to specific types of constraints (e.g., \cite{KetzMcCloskey2023} assume only the sign of parameters is constrained and the critical values for the Likelihood Ratio test of \cite{cavaliere2025uniformcriticalvalueslikelihood} and the bootstrap method of \cite{CAVALIERE2022} assume parameters that may be near the boundary lie in a product space). \cite{FAN2023} provide a method that weakens the product space assumption by allowing the parameter space to be defined by a set of linear inequality constraints. The proximal bootstrap method of \cite{LI2025} allows the parameter space not to be a product space. Also, the proximal bootstrap method does not have the limitation of \cite{KETZ2018}'s method that the initial constrained estimator must be $\sqrt{n}$-consistent. On the other hand, they do not apply their method in cases where the nuisance parameter is high-dimensional, and when the orthogonalized version of the method is used, the $\sqrt{n}$-consistency condition can be weakened for the high-dimensional nuisance parameter.

 Neyman orthogonality (introduced by \cite{Neyman1959}) has been used in many recent papers to achieve asymptotic normality for a low-dimensional parameter of interest after estimation of a high-dimensional nuisance parameter, including \cite{ChernozhukovDoubleDebiased},
\cite{ChernozhukovHansenSpindler2015-General}, \cite{ChernozhukovHansenSpindler2015}, \cite{BelloniChernozhukovHansen2012}, \cite{BelloniChernozhukovHansen2014}, and \cite{belloni2018highdimensional}. Once initial estimates of the parameters have been obtained and the moment conditions have been orthogonalized with respect to the high-dimensional nuisance parameter, these papers obtain an estimate of the parameter of interest either by minimizing the orthogonalized moment conditions directly or by applying a one-step correction to an initial estimate using those conditions. This one-step estimator can be viewed as applying a principle similar to that of the QUE in \cite{KETZ2018}, where the single step is performed using a linearization of the orthogonalized score rather than a linearization of the first-order condition.  In this sense, the orthogonalized QUE is not itself a new approach to estimation. What has been less appreciated is its connection to \cite{KETZ2018} and, more broadly, the benefit this construction provides when the parameter of interest is constrained. While some of these applications allow the high-dimensional nuisance parameter to be constrained, they assume the parameter of interest is interior, and so the debiasing step is understood only as a technique for being insensitive to nuisance parameters rather than for handling the boundary. Yet the same reasoning that shows that \cite{KETZ2018}'s estimator is asymptotically normal at or near the boundary can be applied to the one-step estimate. 
\par 
In Section \ref{s: MandE}, I first introduce the quasi-unconstrained estimator and the high-level conditions for asymptotic normality. I also verify the conditions for a Gaussian random coefficient model with correlated coefficients. Section~\ref{s: InferenceForSubvectors} contains the inference results for the low-dimensional setting, and Section~\ref{s: HighDimensional} extends this to settings with a high-dimensional nuisance parameter when the orthogonalized version of the estimator is used. I verify the high-dimensional version of the conditions in an application involving the estimation of treatment effects with monotonicity constraints and many controls. While controlling size is the central focus of the paper, I also provide counterexamples to the admissibility of tests in Section \ref{s: Power} and discuss why interactive constraints present a challenge to establishing admissibility. I show how the method can be applied in Section~\ref{s: Application} using data from a randomized experiment in \cite{BJORKMAN_GUARISO2027} that studies the effect of educational interventions on child test scores in India. Across the simulation designs fitted to the data of \cite{BJORKMAN_GUARISO2027}, I find that the tests succeed in controlling size. In terms of power, I find that how methods handle the high-dimensional nuisance parameter and whether they use bounds on the parameters of interest are much more important than whether their test statistic makes use of the interactive constraints.



\section{Model and Estimator} \label{s: MandE}

Generally, the parameter to be estimated $\theta$ does not fully specify the distribution of the data, so I use $\omega$ to denote a, usually infinite-dimensional, parameter so that $\gamma := (\theta, \omega)$ fully specifies the distribution of the data. The corresponding parameter space satisfies the conditions given in Assumption \hyperref[A1]{1}.

\begin{assumption}\label{A1} Let $\Gamma = \{ \gamma = (\theta, \omega) \;:\; \theta \in \Theta,\; \omega \in \Omega(\theta)\}$ be compact, where
\begin{enumerate}
    \item $\Theta \subseteq \mathbb{R}^J$ is compact and non-empty.
    \item $\Omega(\theta) \subseteq \Omega$ for all $\theta \in \Theta$ for some compact metric space $\Omega$.
\end{enumerate}
\end{assumption}

Because $\gamma$ specifies the distribution of the data, where I use $n$ to denote the number of observations, it also fully specifies the distribution of the stochastic objective function $\hat{Q}$ being minimized to estimate the value of $\theta$. The true value of $\theta$ is defined to be the unique minimizer over $\Theta$ of the function $Q_n$ to which $\hat{Q}$ converges:

\begin{equation}\label{eq: thetaDefined}
    \theta_n = \argmin_{\theta \in \Theta} Q_n(\theta)
\end{equation}
I use the subscript $n$ to indicate that I consider drifting sequences of the parameter, so $\theta_n$ and $Q_n$ may change with the sample size. In order to use the method of \cite{KETZ2018}, we first need some initial estimate $\hat{\theta}$, which is obtained by minimizing $\hat{Q}$ over $\Theta$. However, it is not necessary for $\hat{\theta}$ to exactly minimize $\hat{Q}$ over $\Theta$, but rather it has to at least approximately achieve the minimum value. More specifically, I assume that
\begin{equation}\label{eq: ConstrainedDefined}
   \hat{Q}  (\hat{\theta}) = \inf_{\theta \in \Theta} \hat{Q} (\theta) + o_p(1/n).
\end{equation}
\par 
\cite{cox2022} characterizes the asymptotic distribution of a constrained extremum estimator under drifting sequences of the parameter $\{ \theta_n \}_{n \in \mathbb{N}}$ with $\theta_n \in \Theta \subseteq \mathbb{R}^J$ in the case where the sequences of sets $\sqrt{n}(\Theta - \theta_n)$ converge to some set $S \subseteq \mathbb{R}^J$. In this case, they show that the results of \cite{Andrews1999} can be generalized so that the asymptotic distribution of $\sqrt{n}(\hat{\theta} - \theta_n)$ is equal to that of a normal random vector projected onto the set $S$. However, the sequence $\sqrt{n}(\Theta - \theta_n)$ will not converge under all drifting sequences $\{ \theta_n \}_{n \in \mathbb{N}}$ so $\sqrt{n}(\hat{\theta} - \theta_n)$ will not always converge in distribution and what it does converge to is significantly different depending on $S$. Because of this, instead of using $\hat{\theta}$ directly for inference, I focus on using the quasi-unconstrained estimator since it is asymptotically normal under quite general conditions on the drifting sequence $\{ \theta_n \}_{n \in \mathbb{N}}$. As a result, I do not rely on $\sqrt{n}(\Theta - \theta_n)$ converging to a particular set and therefore allow for more general conditions on $\Theta$.\footnote{For example, if $\theta_n$ is fixed at some boundary point $\theta^*$, we do not need to impose Chernoff Regularity on $\Theta$ to ensure that $\sqrt{n}(\Theta - \theta_n)$ is converging to the tangent cone of $\Theta$ at $\theta^*$.}
\par 
The quasi-unconstrained estimator is obtained by taking the constrained estimator and performing one Newton-Raphson step using the quadratic approximation of $\hat{Q}$ at $\hat{\theta}$. I use $D Q_n$ and $D^2 Q_n$ to denote the generalized first-order and second-order partial derivatives of $Q_n$, respectively. The partial derivatives of $Q_n$ may not exist for all $\theta \in \Theta$. For example, $Q_n$ and $\hat{Q}$ may not be defined outside $\Theta$, such as in a random coefficient model. In this case, $D Q_n$ and $D^2 Q_n$ would denote the first-order and second-order left or right partial derivatives. Also, $Q_n$ may not be smooth but only “stochastically differentiable"  (see \cite{Pollard1985}), which can be the case in quantile estimation (e.g., \cite{PakesPollard1989}). Also following \cite{KETZ2018}, I allow for the case that these derivatives are not known exactly and must be approximated numerically using $\widehat{D \hat{Q}}$ and $\widehat{D^2 \hat{Q}}$. In Section \ref{s: HighDimensional}, I show that the idea of using approximations of the derivatives can be extended for the orthogonalized version of the method, where part of the "approximation" is using the estimated rather than true values of additional high-dimensional parameters beyond $\theta$. The quasi-unconstrained estimator can then be defined as:
\begin{equation} \label{eq: Quasi-unconstrainedDefined}
    \Tilde{\theta} = \hat{\theta} - (\widehat{D^2 \hat{Q}}(\hat{\theta}))^{-1} \widehat{D \hat{Q}} (\hat{\theta}).
\end{equation}
The compactness conditions in Assumption \hyperref[A1]{1} ensures that for any sequences $\{\gamma_n \}_{n \in \mathbb{N}}$ with $\gamma_n \in \Gamma$, there is a convergent subsequence. I denote these convergent sequences as $\Gamma(\gamma^*) = \{ \{\gamma_n\}_{n \in \mathbb{N}} \;:\; \gamma_n \in \Gamma \text{ for all }n,\; \gamma_n \rightarrow \gamma^* \}$ for some $\gamma^* := (\theta^*, \omega^*) \in \Gamma$. Using the results of \cite{ANDREWS2020}, the asymptotic distribution of the test statistics under these convergent subsequences is used to show that the tests control size uniformly. In order to show the asymptotic behavior of the test statistics under these convergent drifting sequences, I impose conditions in Assumption \hyperref[A2]{2} that guarantee that $\Tilde{\theta}$ is asymptotically normal under these convergent drifting sequences.

\begin{assumption}\label{A2} Under $\{ (\theta_n, \omega_n) \}_{n \in \mathbb{N}} = \{ \gamma_n \}_{n \in \mathbb{N}} \in \Gamma(\gamma^*)$ for any $\gamma^* \in \Gamma$,
\begin{enumerate}
    \item $\sqrt{n}(\hat{\theta} - \theta_n) = O_p(1)$.
    \item $\sqrt{n} D \hat{Q}(\theta_n) \Rightarrow  N(0, V(\gamma^*))$, where $V(\gamma^*)$ is symmetric and positive definite.
    \item $D^2 \hat{Q} (\theta_n) \overset{p}{\rightarrow} \mathcal{J}(\gamma^*)$, where $\mathcal{J}(\gamma^*)$ is symmetric and non-singular.
    \item $|| \widehat{D \hat{Q}}(\hat{\theta}) - D \hat{Q}(\theta_n) - D^2 \hat{Q} (\theta_n)(\hat{\theta} - \theta_n)||_2 = o_p(1/\sqrt{n})$ and \\ $||vec( \widehat{D^2 \hat{Q}}(\hat{\theta})) - vec(D^2 \hat{Q} (\theta_n))||_2 = o_p(1)$.
    \item $\hat{V}(\hat{\theta}) \overset{p}{\rightarrow} V(\gamma^*)$.
\end{enumerate}  
\end{assumption}

Under Assumption \hyperref[A2]{2}, by Theorem 1 of \cite{KETZ2018}, under $\{ \gamma_n \}_{n \in \mathbb{N}} \in \Gamma(\gamma^*)$, $$\sqrt{n}(\Tilde{\theta} - \theta_n) \Rightarrow  \mathcal{Z} \sim N(0_J, \Sigma(\gamma^*))$$ where $\Sigma(\gamma^*) = \mathcal{J}(\gamma^*)^{-1}V(\gamma^*)\mathcal{J}(\gamma^*)^{-1}$ and $\hat{\Sigma} = (\widehat{D^2 \hat{Q}}(\hat{\theta)})^{-1} \hat{V}(\hat{\theta})(\widehat{D^2 \hat{Q}}(\hat{\theta}))^{-1} \overset{p}{\rightarrow} \Sigma(\gamma^*)$. In other words, $\Tilde{\theta}$ is asymptotically normal and its asymptotic variance can be consistently estimated.\footnote{For notational simplicity, I suppress the dependence of $\Sigma$ on $\gamma^*$ below.}
\par 
I show in Appendix B in the Supplementary Materials how Assumption \hyperref[A2]{2.2} is satisfied when $\sqrt{n}(D\hat{Q} - D Q_n)$ converges weakly to a Gaussian process on $\Theta$. Since $D \hat{Q} (\theta)$ often involves a sample average, in many cases this can be shown by applying the proper Uniform Central Limit Theorem. The asymptotic normality condition in Assumption \hyperref[A2]{2.2} provides the basis for the asymptotic normality of $\Tilde{\theta}$, because the quasi-unconstrained estimator will asymptotically satisfy a first-order condition. In order for Assumption \hyperref[A2]{2.3} to hold, we want that $\sup_{\theta \in \Theta} ||vec(D^2 \hat{Q} (\theta)) - vec(D^2 Q_n(\theta))||_2 \overset{p}{\rightarrow} 0$. If $D \hat{Q}$ and $D^2 \hat{Q}$ must be approximated numerically, Assumption \hyperref[A2]{2.4} requires that these approximations converge to the actual first and second generalized derivatives at a sufficiently fast rate as the sample size grows. Assumption \hyperref[A2]{2.5} simply requires that the variance of the derivative at $\theta_n$ can be consistently estimated using $\hat{\theta}$. Assumption \hyperref[A2]{2.1} imposes that $\hat{\theta}$ is $\sqrt{n}$-consistent without requiring that it is asymptotically normal. I also show in Appendix B that when the other conditions in Assumption \hyperref[A2]{2} hold, only one additional weaker condition is needed to show that any consistent estimator is also $\sqrt{n}$-consistent.

\subsection{Correlated Random Coefficient Application}
\label{subsec:grc}

I now show how the conditions of Assumptions \hyperref[A1]{1} and \hyperref[A2]{2} can be verified in a model with correlated random coefficients. In the common Gaussian model where the random coefficients are normally distributed with an unknown covariance matrix $\Omega$, if the random coefficients are uncorrelated, then the only constraints are nonnegativity constraints on the variances, which fits with the product space assumption of \cite{KETZ2018}, while if the random coefficients are correlated, then $\Omega$ is constrained to be positive semidefinite, which means interactive constraints on the elements of $\Omega$. Suppose we have clustered observations where the number of independent clusters, $G$, replaces $n$ in Assumption \hyperref[A2]{2}. We observe $G$ independent clusters $W_g=(Y_g,X_g,Z_g)$ with $m$ observations per cluster, with $Y_g\in\mathbb R^m$, $X_g\in\mathbb R^{m\times p}$, and $Z_g\in\mathbb R^{m\times2}$, where $m,p$ are fixed. The model is
\begin{equation}\label{grc:model}
 Y_g=X_g\beta+Z_gb_g+\varepsilon_g,\qquad
 b_g\sim N(0,\Omega),\quad \varepsilon_g\sim N(0,\tau I_m).
\end{equation}
If we want to allow for the case where $Z_g$ includes a constant, so one of the elements of $b_g$ serves as a random intercept, it would not be possible to separate this intercept from the error term if we only had one observation per cluster. Here, since the error terms $\varepsilon$ are uncorrelated within each cluster and the random coefficients vary at the cluster level, data with a few observations per cluster can be used to identify the variance of a random-intercept separately from the variance of the error term, since $\varepsilon$ but not the random coefficients contributes to the variation in the outcome within each cluster.
We can write
\[
 \Omega=\begin{pmatrix}u&w\\w&v\end{pmatrix},\quad
 \xi=(\tau,u,v,w)',\quad \theta=(\beta',\xi')',\quad J=p+4.
\]
Fix a nonempty compact convex set $\mathcal B\subset\mathbb R^p$ and constants $0<\underline\tau<\overline\tau<\infty$, and $B>0$. Constraining $\Omega$ to be positive semidefinite means $u\ge0$, $v\ge0$, and $w^2\le uv$. Therefore, define
\[
 \Theta=\mathcal B\times[\underline\tau,\overline\tau]
 \times\{(u,v,w) \in [-B,B]^3: 0\le u, 0 \le v , \; w^2 \le uv  \}.
\]
Let $\mathcal K\subset\mathbb R^{m\times p}\times\mathbb R^{m\times2}$ be a compact support of the regressor pair and $\mathcal F$ be the set of the probability measures on $\mathcal K$ satisfying Condition~\ref{grc:primitive}(i) below, so the space for $\gamma$ will be given by $\Gamma=\Theta\times\mathcal F$. Let the criterion function be proportional to the average negative Gaussian log-likelihood, so $\hat Q(\theta)=\frac1G\sum_{g=1}^G q(W_g;\theta)$ and  $Q_\gamma(\theta)=\mathbb E_\gamma [ q(W;\theta)]$ with
$$q(W;\theta)=\frac12\log\det C_Z(\theta) +\frac12(Y-X\beta)'C_Z(\theta)^{-1}(Y-X\beta),$$
where $C_Z(\theta)=\tau I_m+Z\Omega Z'$ captures the variance of $Y_g$ conditional on the regressors $(X_g,Z_g)$.
\begin{condition}\label{grc:primitive}
The model class and estimator defined above satisfy the following.
\begin{enumerate}[label=(\roman*)]
\item For fixed $c_X,c_C>0$, every $F\in\mathcal F$ satisfies $ \mathbb E_F[X'X]\succeq c_X I_p$ and
\begin{equation*}
    \mathbb E_F [\norm{vec(aI_m+ZHZ')}_2^2] \ge c_C(a^2+\norm{vec(H)}_2^2)
\text{ for all }a\in\mathbb R,\ H\in\mathbb S^2.
\end{equation*}
\item Under each $\gamma\in\Gamma$, the model in equation~\eqref{grc:model} holds, clusters are i.i.d. with $(X_g,Z_g)$ having the joint distribution $F$, and $b_g$ and $\varepsilon_g$ are mutually independent and independent of $(X_g,Z_g)$. 
\item Let $\hat\theta\in\Theta$ be an approximate minimizer satisfying
\begin{equation*}\label{grc:optimization}
 \hat Q(\hat\theta)\le\inf_{\theta\in\Theta}\hat Q(\theta)+ r_G, 
\end{equation*}
where $r_G = o_p(1/G)$, along every $\{\gamma_G\}_{G \in \mathbb{N}} \in \Gamma(\gamma^*)$ for any $\gamma^* \in \Gamma$.
\end{enumerate}
\end{condition}


Part (i) imposes strong identification of the mean and covariance parameters. The first inequality requires enough variation in $X$ to distinguish every direction in $\beta$. In the second, $a$ is a change in $\tau$ and $H\in\mathbb S^2$ is a change in $\Omega$, where $\mathbb S^2$ denotes all real symmetric $2\times2$ matrices. Their combined effect on the conditional covariance of $Y$ is $aI_m+ZHZ'$. The lower bound requires every nonzero pair $(a,H)$ to produce a change on average across designs. 

\begin{proposition}\label{grc:verification}
Under Condition~\ref{grc:primitive}, Assumptions~\ref{A1} and~\ref{A2}
hold for $\hat Q$ and $\hat\theta$ using exact derivatives. More specifically, let $s(W;\theta)=Dq(W;\theta)$ be the gradient of the negative conditional log-likelihood for one cluster. Along every sequence $\{\gamma_G\}_{G\in\mathbb N}\in\Gamma(\gamma^*)$,
$\widetilde\theta=\hat\theta- D^2\hat Q(\hat\theta)^{-1}D\hat Q(\hat\theta)$ satisfies
\[
 \sqrt G(\widetilde\theta-\theta_G)
 \Rightarrow N\bigl(0,\Sigma(\gamma^*)\bigr),
\]
where $\Sigma(\gamma)= \mathbb E_{\gamma}[D^2 q(W;\theta)]^{-1} = \mathbb E_{\gamma} [s(W;\theta)s(W;\theta)']$. Furthermore, \\$\hat\Sigma= D^2\hat Q(\hat\theta)^{-1}\hat V(\hat\theta) D^2\hat Q(\hat\theta)^{-1} \overset{p}{\rightarrow} \Sigma(\gamma^*)$, where
$\hat V(\theta)=G^{-1}\sum_{g=1}^G s(W_g;\theta)s(W_g;\theta)'$.
\end{proposition}

Placing a lower bound on the variance of the error term $\varepsilon$ ensures a smooth observed likelihood even when $\Omega$ is singular, and the restrictions in Condition~\ref{grc:primitive}(i) ensure strong identification and that the information matrix isn't close to being singular. Note that this is distinct from requiring the eigenvalues of $\Omega$ to be bounded away from zero, and these drifting sequences include cases where $\Omega_G$ is singular or approaches singularity. In Appendix B of the Supplementary Materials, I give a concrete example for these strong identification restrictions. Other conditions are used for convenience but could be weakened. For example, uniformly bounded variable cluster sizes and any fixed number of random coefficients could likely be accommodated with adjustments to the identification conditions. Extensions could also be made for when the likelihood is misspecified and $\theta_G$ denotes the pseudo-true values of the parameters, with the limiting variance then generally being a sandwich matrix.

\section{Low-Dimensional Setting}\label{s: InferenceForSubvectors}



I now show that tests using this quasi-unconstrained estimator can control the size uniformly over $\Theta$ asymptotically. Let $CP_n(\gamma)$ denote the coverage probability for $\theta$ of a given test when the sample size is $n$ (i.e., the probability of failing to reject the null hypothesis that $\theta_n = \theta$). Following \cite{ANDREWS2020}, let $$AsySz = \liminf_{n \rightarrow \infty} \inf_{\gamma \in \Gamma} CP_n(\gamma)$$ be the asymptotic size and let $$AsyMaxCP = \limsup_{n \rightarrow \infty} \sup_{\gamma \in \Gamma} CP_n(\gamma)$$ be the asymptotic maximum coverage probability. When the nominal size of a test is equal to $\alpha$, a test provides asymptotic size control in a uniform sense when $AsySz \geq 1 - \alpha$. In this case, we can be confident that the test does not over-reject when the sample size is large, even when the parameter is on or near the boundary. Uniform size control is shown in the proofs using the general technique of \cite{ANDREWS2020} of showing that for any sequence $\{\gamma_n\}_{n \in \mathbb{N}}$, there is a subsequence under which size is being controlled. \cite{KETZ2018} also uses the technique of \cite{ANDREWS2020}, but focuses specifically on when $\Theta$ is a product space. When $\Theta$ is a product space, there are no interactions between when one element is close to the boundary and another element is. As a result, the problem of considering all sequences where $\theta_n$ converges to a point on the boundary can be reduced to considering only a finite number of cases based on whether each element of $\theta$ is converging to one of its bounds. The key additional insight used in the proof is that because $\Theta \subset \mathbb{R}^J$, the sequence of sets $\sqrt{n}(\Theta - \theta_n)$ must have a convergent subsequence, and I am able to focus on analyzing the asymptotic distribution of the test statistic under these subsequences.

I focus on the case in which the null hypothesis specifies 
the value of a subvector of $\theta$. Following \cite{KETZ2018}, let
$\theta=(\beta',\delta')'$ where $\beta\in\mathbb R^K$ is the parameter
of interest and $\delta\in\mathbb R^L$ is a nuisance parameter. Let
\begin{equation*}
 \begin{bmatrix}
  \Sigma_{\beta\beta}&\Sigma_{\beta\delta}\\
  \Sigma_{\delta\beta}&\Sigma_{\delta\delta}
 \end{bmatrix}
\end{equation*}
denote the conformable submatrices of $\Sigma$. We are interested in
testing
\begin{equation*}
 H_0:\ \beta=\beta_0,\: \delta\in D_{\beta_0}
 \quad vs.\quad
 H_1:\ \beta\ne\beta_0,\: \theta \in \Theta
\end{equation*}
where $D_{\beta_0}:=\{\delta\in\mathbb R^L\mid
(\beta_0,\delta)\in\Theta\}$. For the special case considered by \cite{KETZ2018} where $\Theta$
is equal to the Cartesian product of closed, possibly infinite,
intervals, there is no interaction between the true value of the
subvector of interest and the feasible set for the nuisance
parameter. However, in general $D_{\beta_0}$ depends on $\beta_0$.
\par
The results from the Wald statistics can be directly extended to
this case, where the test statistic is now equal to
\begin{equation*}
 (\Tilde{\beta}-\beta_0)'(\hat\Sigma_{\beta\beta}/n)^{-1}
       (\Tilde{\beta}-\beta_0).
\end{equation*}
This is because the pointwise asymptotic distribution of the Wald
statistic for $\beta$ does not depend on $\delta$ and is therefore
still fully specified under the null hypothesis. However, for the Likelihood Ratio and Lagrange Multiplier tests, the true value of $\delta$ does affect the asymptotic behavior of the test statistic. Luckily, this can be addressed using the conditioning principle of \cite{Moreira2003}. Under the null hypothesis, one can use
$X_n:=\Tilde{\delta}-\hat\Sigma_{\delta\beta}
\hat\Sigma_{\beta\beta}^{-1}\Tilde{\beta}$ as an asymptotically
sufficient statistic for the unknown nuisance parameter $\delta$.
In the limiting Gaussian experiment, the corresponding centered and
scaled subvectors are independent, so conditioning on the nuisance
statistic gives a nuisance-parameter-free Gaussian reference law.
\par
The Lagrange Multiplier test statistic is typically defined using the derivative of the log-likelihood. Here, if we use a likelihood
function based on the Gaussian approximation to $\Tilde{\theta}$, and the corresponding quadratic statistic minimizes over $D_{\beta_0}$, this gives
\begin{equation*}
     \inf_{d \in D_{\beta_0}} \begin{pmatrix}
    \Tilde{\beta} - \beta_0 \\
    \Tilde{\delta} - d
    \end{pmatrix}'
    (\hat{\Sigma}/n)^{-1}
        \begin{pmatrix}
    \Tilde{\beta} - \beta_0 \\
    \Tilde{\delta} - d
    \end{pmatrix}.
\end{equation*}
Using $\Tilde{\delta}=X_n+\hat\Sigma_{\delta\beta}
\hat\Sigma_{\beta\beta}^{-1}\Tilde{\beta}$, we can express its
conditional version as
\begin{equation*}
    CLM(\beta_0, \Tilde{\beta}, \hat{\Sigma}/n, X_n) =   \inf_{d \in D_{\beta_0}}
    \begin{pmatrix}
    \Tilde{\beta} - \beta_0 \\
    X_n + \hat{\Sigma}_{\delta \beta} \hat{\Sigma}_{\beta \beta}^{-1}\Tilde{\beta} - d
    \end{pmatrix}'
    (\hat{\Sigma}/n)^{-1}
        \begin{pmatrix}
       \Tilde{\beta} - \beta_0 \\
    X_n + \hat{\Sigma}_{\delta \beta} \hat{\Sigma}_{\beta \beta}^{-1}\Tilde{\beta} - d
    \end{pmatrix}.
\end{equation*}
The CLR statistic subtracts the corresponding minimum over the full
parameter space:
\begin{equation*}
\begin{split}
 CLR(\beta_0,\Tilde{\beta},\hat\Sigma/n,X_n)
 ={}&CLM(\beta_0,\Tilde{\beta},\hat\Sigma/n,X_n)\\
 &-\inf_{(b,d)\in\Theta}
 \begin{pmatrix}
  \Tilde{\beta}-b\\
  X_n+\hat\Sigma_{\delta\beta}\hat\Sigma_{\beta\beta}^{-1}
             \Tilde{\beta}-d
 \end{pmatrix}'
 (\hat\Sigma/n)^{-1}
 \begin{pmatrix}
  \Tilde{\beta}-b\\
  X_n+\hat\Sigma_{\delta\beta}\hat\Sigma_{\beta\beta}^{-1}
             \Tilde{\beta}-d
 \end{pmatrix}.
\end{split}
\end{equation*}
 Note that here the likelihood is based on the asymptotic distribution of $\Tilde{\theta}$ rather than the likelihood of the data. This might reasonably be thought of as a type of quasi-likelihood ratio test, but here I simply refer to it as the CLR to be consistent with \cite{KETZ2018}, and I follow the analogous naming convention for the CLM.
\par

Since CLM statistic is equal to the first term of the CLR
statistic, when $\Theta = \mathbb{R}^J$ all three tests are equivalent.\footnote{This is unsurprising given that these three test statistics are known to be identical when the log-likelihood is simply a quadratic function of $\theta$ and $\theta$ is unconstrained (see, e.g., Lemma 1 of \cite{ENGLE1984}).} One difference between my CLR test statistic and the CLR test used by \cite{KETZ2018}, is the feasible set. Since they assume $\Theta$ is the Cartesian product of intervals of the form $\theta_j \in [0,c]$ or $\theta_j \in [-c,c]$ for some large constant $c$, they set the feasible set in the CLR statistic to be equal to the Cartesian product of the extensions of these intervals (i.e., $\theta_j \in [0,\infty)$ or $\theta_j \in (-\infty, \infty)$). They do this because it makes sure that only boundaries at zero are taken into account, and they focus on drifting sequences where elements of $\theta_n$ may be drifting towards zero. In my approach, I use $\Theta$ as the feasible set to be agnostic about which part of the boundary the parameter is close to. $\Theta$ is used not only in calculating the test statistic, but also in calculating the critical values. For the Wald test, calculating p-values and confidence intervals is straightforward because the critical values are always calculated using a chi-squared distribution. The critical values for the CLM and CLR tests are based on a normal distribution projected onto a set which now depends on $X_n$. More precisely, the critical values come from $CLR(\beta_0, n^{-1/2}\hat{\mathcal{Z}}_b + \beta_0, \hat{\Sigma}/n, X_n)$ and $CLM(\beta_0, n^{-1/2}\hat{\mathcal{Z}}_b + \beta_0, \hat{\Sigma}/n, X_n)$, where $\hat{\mathcal{Z}}_b \sim N (0_K, \hat{\Sigma}_{\beta \beta})| X_n, \hat{\Sigma}$. So, for example, for the CLR the critical value at the $1- \alpha$ level is given by
\begin{equation*}
    cv_{1-\alpha}(CLR(\beta_0, n^{-1/2}\hat{\mathcal{Z}}_b + \beta_0,\Sigma, x)|x = X_n,\Sigma = \hat{\Sigma}/n) 
\end{equation*}
\begin{equation*}
    = \inf \{q \in \mathbb{R}\;:\: P(CLR(\beta_0, n^{-1/2}\hat{\mathcal{Z}}_b+\beta_0,\Sigma,  x) \leq q \;|\;x = X_n, \Sigma = \hat{\Sigma}/n) \geq 1 - \alpha\}.
\end{equation*}
Generally, critical values must be calculated numerically by repeatedly sampling from $\hat{\mathcal{Z}}_{b}$. When $\theta$ is low-dimensional, this is generally not computationally intensive, although the critical values depend on the null value, so this must be done repeatedly to invert the test for confidence sets. I show in Appendix C of the Supplementary Materials that, depending on the form of the parameter space $\Theta$, we can increase the computational speed of projecting $n^{-1/2}\hat{\mathcal{Z}}_b + \beta_0$ using a change of variable and approximations for the new feasible set after that change of variable. When the full vector is high-dimensional, but the parameter of interest is low-dimensional, one extra benefit of using the orthogonalized version of this method is that it helps retain computational tractability for conducting inference, as the resampling and projecting need to be done in a low-dimensional space.

\begin{proposition}[Conditional Lagrange Multiplier and subvector Wald]
\label{P4}
If Assumptions~\ref{A1} and~\ref{A2} hold, then the CLM test using
\begin{equation*}
 cv_{1-\alpha}\!\left(
 CLM(\beta_0,n^{-1/2}\hat{\mathcal Z}_b+\beta_0,
          \hat\Sigma/n,X_n)\right)
\end{equation*}
and the subvector Wald test using $cv_{1-\alpha}(\chi_K^2)$ are
uniformly asymptotically equivalent. Both satisfy the uniform
similarity condition $AsySz=AsyMaxCP=1-\alpha$.
\end{proposition}

Proposition~\ref{P4} follows from completing the square:
\begin{equation*}
 CLM(\beta_0,\Tilde{\beta},\hat\Sigma/n,X_n)
 =(\Tilde{\beta}-\beta_0)'(\hat\Sigma_{\beta\beta}/n)^{-1}
       (\Tilde{\beta}-\beta_0)+a_n(\beta_0),
\end{equation*}
where $a_n(\beta_0)$ is the minimized nuisance quadratic, fixed
conditional on $(X_n,\hat\Sigma)$. The exact conditional Gaussian
reference distribution is therefore $\chi_K^2+a_n(\beta_0)$, so the
CLM critical value is $cv_{1-\alpha}(\chi_K^2)+a_n(\beta_0)$.
The same additive term cancels from the rejection rule for every
feasible null subvector, including boundary points. The Wald
statistic uses the quasi-unconstrained estimator $\Tilde{\beta}$,
whose Gaussian limit holds along drifting sequences even at the
boundary. Consequently, both tests satisfy uniform asymptotic
similarity. This extends the conditional score/Wald equivalence
noted in \cite[Remark~1]{KETZ2018} to the parameter spaces considered
here. The cancellation does not require $\Theta$ to be a product
space. Because of this equivalence, I focus on comparing the Wald and CLR tests throughout the remainder of the paper.

\begin{proposition}[Conditional Likelihood Ratio]\label{P3}
If Assumptions~\ref{A1} and~\ref{A2} hold, then, for any fixed
$\epsilon>0$, the CLR test that rejects $\beta=\beta_0$ when
$$ CLR(\beta_0,\Tilde{\beta},\hat\Sigma/n,X_n)
 >cv_{1-\alpha}\!\left(
 CLR(\beta_0,n^{-1/2}\hat{\mathcal Z}_b+\beta_0,
          \hat\Sigma/n,X_n)\right)+\epsilon$$
satisfies $AsySz\ge1-\alpha$.
\end{proposition}

While the Wald and CLM tests satisfy uniform asymptotic similarity,
the stronger condition $AsySz=AsyMaxCP=1-\alpha$ does not hold for
the CLR test in general. Because its reference distribution involves
projecting a normal random vector onto a set, there may be
discontinuities in that distribution, including an atom at the
conditional critical value. As a result, it may not be possible to
choose the threshold so that the conditional probability of
exceeding it is exactly equal to $\alpha$. This is related to the fact that the Wald test is pivotal, but the CLR test is not. For the Wald test, its critical values always come from a chi-squared distribution so they always come from a continuous distribution, but the reference distribution for the CLR does have discontinuities in some cases. For example, there may be a positive probability of the two terms that compose the CLR statistic being equal to each other, so the distribution function does have a discontinuity at zero.\footnote{This is true in a case discussed by \cite{KETZ2018} where $\Theta = [0, \infty)^J$ and $\theta$ is equal to the zero vector.}
\par
 The fixed $\epsilon>0$ in Proposition~\ref{P3} is also added to deal with an issue that arises due to the potential discontinuity. A critical value that controls size in the reference experiment may fail to do so asymptotically with the sampling distribution because weak convergence need not preserve acceptance probabilities when arbitrarily small approximation errors can change decisions on the limiting event. Adding a small positive constant gives the required lower bound on coverage without a continuity assumption at the critical value and without requiring the coverage probabilities to converge at the price of making the test slightly more conservative. The correction follows the approach of \cite[Theorem~1]{AndrewsMikusheva2016}, who establish uniform size control for conditional tests with an arbitrarily small fixed positive addition to the critical value. A related correction in \cite{AndrewsShi2013} adds a fixed positive constant and also increases the quantile level, and they suggest $10^{-6}$ as an example and report identical simulation results with this value and zero. Following this guidance, a small constant such as $\epsilon=10^{-6}$ can be specified in advance, with sensitivity to larger fixed values reported as a practical check.

\section{High-Dimensional Setting}\label{s: HighDimensional}

All results presented so far, as well as those in \cite{KETZ2018}, assume that the dimension of the estimated parameters is fixed, so they may only provide a good approximation in finite samples where the number of estimated parameters is small relative to $n$. However, it is not uncommon to have applications in which the parameter vector is both constrained and high-dimensional. For example, the control weights for a synthetic control unit are usually constrained and often have a dimension comparable to the number of time periods. Also, some applications of the random coefficient model of \cite{BLP} have many markets or many products so $\theta$ is both high-dimensional and has some of its elements constrained (see, for example, \cite{Armstrong2016}). I explore another example of where this arises in Subsection~\ref{sub: MontoneTreatment} below.
\par 
The strategy in this Section is to first estimate the high-dimensional nuisance parameter, possibly using regularization, and then estimate the remaining parameters using a Neyman orthogonalized score. The QUE is then estimated by linearizing this orthogonalized score, in the same way it was estimated earlier by linearizing the first-order condition. To make that correspondence transparent, it is convenient to slightly change the notation. From this point forward, $\theta \in \Theta \subseteq \mathbb{R}^{J}$ denotes only the parameters we do not orthogonalize with respect to, and $\psi \in \Psi_n \subseteq \mathbb{R}^{P_n}$ denotes the high-dimensional nuisance parameter we do, with $J$ fixed and $P_n$ permitted to grow with $n$. The criterion function before orthogonalization, therefore, depends on the pair $(\theta,\psi)$ and we can view $\psi$ as being a component of the infinite-dimensional nuisance parameter $\omega$.\footnote{Note that $\psi$ and the nuisance subvector $\delta$ of Section~\ref{s: InferenceForSubvectors} are distinct objects handled by different devices, orthogonalization in the former case and conditioning in the latter.} After orthogonalizing, the resulting moment conditions are a function of $\theta$ alone. 
\par 
Suppressing its dependence on $n$, the population orthogonalized score can be defined as
\begin{equation}\label{Neyman Orthogonal Score}
    M(\theta, \psi, \eta) = D_\theta Q_n(\theta,\psi) - \eta D_\psi Q_n(\theta,\psi),
\end{equation}
where $\eta \in \mathbb{R}^{J \times P_n}$ is an additional nuisance parameter and the true value of this matrix $\eta_n$ satisfies $$  D_\psi M(\theta_n, \psi_n, \eta_n) = 0_{J \times P_n}.$$
Using some estimate of this matrix $\hat{\eta}$, we can define the sample orthogonalized score as
\begin{equation}\label{Sample Neyman Orthogonal Score}
    \hat{M}(\theta,\hat{\psi},\hat{\eta}) = \widehat{D_\theta \hat{Q}}(\theta,\hat{\psi}) - \hat{\eta} \widehat{D_\psi \hat{Q}}(\theta,\hat{\psi}).
\end{equation}
Under suitable rates of convergence conditions on $\hat{\eta}$ and $\hat{\psi}$ and smoothness conditions on $Q_n$, an estimator using the moment conditions $\hat{M}(\theta,\hat{\psi},\hat{\eta})$ can be asymptotically equivalent to one using $\hat{M}(\theta,\psi_n,\eta_n)$. 
\par
We can now define the constrained estimator $\hat{\theta}$ using a GMM criterion function:

\begin{equation}\label{eq: GMM Estimator}
   \hat{\theta} = \argmin_{\theta \in \Theta}  \hat{M}(\theta,\hat{\psi},\hat{\eta})'\hat{M}(\theta,\hat{\psi},\hat{\eta}).
\end{equation}

In the low-dimensional setting, the QUE solves the
linearized first-order condition
$\widehat{D \hat{Q}}(\hat\theta)
 + \widehat{D^2 \hat{Q}}(\hat\theta)
 (\theta-\hat\theta) = 0$. In the orthogonalized setting, the linearization $ \hat M(\hat\theta,\hat\psi,\hat\eta)
 + \partial_\theta\hat M(\hat\theta,\hat\psi,\hat\eta)(\theta-\hat\theta) = 0$ is used. Therefore, the quasi-unconstrained estimator is now equal to
\begin{equation}
    \Tilde{\theta} = \hat{\theta} - (\partial_\theta \hat{M}(\hat{\theta},\hat{\psi},\hat{\eta})'   \partial_\theta \hat{M}(\hat{\theta},\hat{\psi},\hat{\eta}))^{-1}\partial_\theta \hat{M}(\hat{\theta},\hat{\psi},\hat{\eta})'  \hat{M}(\hat{\theta},\hat{\psi},\hat{\eta}).
\end{equation}
\par 
Note that in the special case where $\hat{M}(\theta,\hat{\psi},\hat{\eta})$ is affine in $\theta$, this is equivalent to setting the orthogonalized moment conditions equal to zero,\footnote{This estimator can also be viewed as performing a Gauss-Newton update of the GMM objective function at the initial estimate $\hat{\theta}$.} so $\Tilde{\theta}$ is defined by
$$ \hat{M}(\Tilde{\theta},\hat{\psi},\hat{\eta}) = 0.$$

\begin{assumption}\label{A3}
Along every sequence
$\{\gamma_n\}_{n\in\mathbb N}=\{(\theta_n,\omega_n)\}_{n\in\mathbb N}
\in\Gamma(\gamma^*)$ with $\gamma^* \in \Gamma$, write
$\hat M_\theta=\partial_\theta\hat M(\hat\theta,\hat\psi,\hat\eta)$.
The following hold.
\begin{enumerate}
\item $\sqrt n(\hat\theta-\theta_n)=O_p(1)$.

\item $\hat M(\hat\theta,\hat\psi,\hat\eta)
 -\hat M(\theta_n,\psi_n,\eta_n)
 =\partial_\theta\hat M(\theta_n,\psi_n,\eta_n)
       (\hat\theta-\theta_n)+o_p(n^{-1/2})$ \\ and $ \hat M_\theta-\partial_\theta\hat M(\theta_n,\psi_n,\eta_n)
 =o_p(1).$

\item $\sqrt n\,\hat M(\theta_n,\psi_n,\eta_n)
\Rightarrow N(0_J,V_M(\gamma^*))$ for a positive
definite matrix $V_M(\gamma^*)$.

\item $\partial_\theta\hat M(\theta_n,\psi_n,\eta_n)
\overset{p}{\rightarrow}M_\theta$, where $M_\theta$ has column
rank $J$ and may depend on $\gamma^*$.

\item $\hat V_M(\hat\theta)\overset{p}{\rightarrow}
V_M(\gamma^*)$.
\end{enumerate}
\end{assumption}

Assumption~\ref{A3} imposes conditions exactly analogous to those imposed by Assumption~\ref{A2}. In particular, Assumption \hyperref[A3]{3.2} is the analog of Assumption \hyperref[A2]{2.4}, where we can view $\theta$ as the parameter vector and view the error in the estimated nuisance parameters $\hat{\psi}$ and $\hat{\eta}$ as part of the numerical approximation error in estimating the first and second derivatives of the criterion function. Because $\hat{\theta}$ is $\sqrt{n}$-consistent, the asymptotic distribution of $\tilde{\theta}$ will depend on the limiting behavior of $\sqrt{n}\hat{M}(\theta_n,\hat{\psi},\hat{\eta})$ and
$\partial_{\theta}\hat{M}(\theta_n,\hat{\psi},\hat{\eta})$, and what we require is that these be asymptotically equivalent to $\sqrt{n}\hat{M}(\theta_n,\psi_n,\eta_n)$ and
$\partial_{\theta}\hat{M}(\theta_n,\psi_n,\eta_n)$. The rates required are the same as in Assumption \hyperref[A2]{2.4} since the score must agree to order $o_p(1/\sqrt{n})$, while the derivative of the score needs only to agree to order $o_p(1)$, which Assumption \hyperref[A3]{3.2} imposes.

\begin{proposition}[Orthogonalized Estimator]\label{P5}
Suppose Assumptions~\ref{A1} and~\ref{A3} hold and, along each
$\{\gamma_n\}_{n\in\mathbb N}=\{(\theta_n,\omega_n)\}_{n\in\mathbb N}
\in\Gamma(\gamma^*)$, and $M(\theta_n,\psi_n,\eta_n) = 0$ holds for every $n$. Define
\[
 \hat\Sigma
 =(\hat M_\theta'\hat M_\theta)^{-1}
   \hat M_\theta'\hat V_M(\hat\theta)\hat M_\theta
   (\hat M_\theta'\hat M_\theta)^{-1}.
\]
Then, along every such sequence, as $n\to\infty$ with $J$ fixed and
$P_n\to\infty$,
\[
 \sqrt n(\widetilde\theta-\theta_n)\Rightarrow 
 N(0,\Sigma(\gamma^*)),
 \qquad
 \hat\Sigma\overset{p}{\rightarrow}\Sigma(\gamma^*),
\]
where $\Sigma(\gamma^*)=(M_\theta'M_\theta)^{-1}M_\theta'V_M(\gamma^*)
M_\theta(M_\theta'M_\theta)^{-1}$.
\end{proposition}

The earlier results then apply with $\theta$ read as the full parameter vector, with $\psi_n$ and $\eta_n$ absorbed into the infinite-dimensional component $\omega_n$ of $\gamma_n$, so that Assumption \hyperref[A1]{1} and the convergent subsequences $\Gamma(\gamma^{*})$ retain their meaning without modification. The proof follows similar reasoning as Theorem 1 of \cite{KETZ2018}. As a result, size can be controlled in a uniform sense using the methods in Section \ref{s: InferenceForSubvectors}. However, I have implicitly assumed in this Section that the parameter space for $\theta$ does not depend on $\psi$, which is an important caveat for using the CLR test in this setting.


\begin{remark}[Interactive Constraints Between High- and Low-dimension Parameters]\label{r: interactionHighDim}
    One complication of the orthogonalization approach that the conditional approach of Section~\ref{s: InferenceForSubvectors} does not share is accommodating constraints involving both $\psi$ and $\theta$. In the CLR statistic, the first term is minimized over all $\delta \in D_{\beta_0}$ and the second over all $\theta \in \Theta$, so the statistic is computed without imposing any restriction on how the constraints on the parameter of interest and those on the nuisance parameter interact. Here, by not having $\Theta$ depend on $\psi$, we are not allowing interactive constraints as we were when allowing $D$ to depend on $\beta$. Because we do not want to rely on $\sqrt{n}$-consistency and asymptotic normality of the estimate of the $P_n$-dimensional vector $\psi$, and because minimizing over a $P_n$-dimensional set is computationally burdensome, we now form the likelihood using only $\tilde{\theta}$ and $\hat{\Sigma}$. The feasible set entering the CLR statistic is then a set for $\theta$ alone. Writing $\mathcal{S}_n \subseteq \mathbb{R}^{J} \times \mathbb{R}^{P_n}$ for the joint constraint set on $(\theta,\psi)$, the relevant set is $ \Theta_{\psi_n} := \{\theta \in \mathbb{R}^{J} : (\theta,\psi_n) \in
    \mathcal{S}_n\},$ which depends on the unknown $\psi_n$. For the statistic and its critical values to use the actual parameter space for $\theta$, the feasible set for $\theta$ must therefore not vary with the value of the nuisance parameter, which is to say: $ \mathcal{S}_n = \Theta \times \Psi_n.$ When there are interactions between the constraints on $\theta$ and $\psi$, one could alternatively define the CLR statistic using $\Theta = \{ \theta : (\theta,\psi) \in \mathcal{S}_n \text{ for some } \psi \text{ and } n\}$. Provided Assumptions \hyperref[A1]{1} and \hyperref[A3]{3} hold for this $\Theta$, we could control size uniformly over $\Theta$ and hence uniformly over $\Theta_{\psi_n}$. In such cases, the potential cost of using the orthogonalized version of the test would not be in losing the ability to control size, but in not being able to make use of restrictions specific to $\psi$ to achieve higher power.
\end{remark}

\subsection{Monotone Treatment with Many Controls Application}\label{sub: MontoneTreatment}

A common context in which there is a low-dimensional parameter of interest and a high-dimensional nuisance parameter is estimation of treatment effects with many control variables. If there are constraints on the relationship between these treatment effects, then both interactive constraints and high dimensionality arise. To illustrate this, I verify Assumptions 1 and 3 using an example in which both treatment effects are subject to monotonicity constraints and there are many controls. Consider two binary interventions that individuals may receive one, both, or neither of, with treatment-arm indicators $D=(D_{10},D_{01},D_{11})'$ and the untreated arm omitted. If the marginal effect of each intervention is nonnegative, we may impose monotonicity on the three regression effects,

\[
 \Theta=\{\theta\in[0,B]^3: \theta_{10}\le\theta_{11},\ \theta_{01}\le\theta_{11}\},\qquad B>0.
\]
The restrictions impose that each intervention has a nonnegative effect on its own and a nonnegative incremental effect, but do not impose that the effects of the interventions are additive. Treatment assignment may depend on a growing collection of controls. For each $n$, observe $\{(Y_i,D_i,X_i^{[P_n]})\}_{i=1}^n$, where $P_n\to\infty$, $X^{[p]}=(X_1,\ldots,X_p)'$, and
\begin{equation}\label{mt:model}
 Y_i=D_i'\theta+X_i^{[P_n]\prime}\psi^{[P_n]}+\varepsilon_i,
 \qquad \Pr(D_i=\mathbf e_a\mid X_i)=e_{a,n}(X_i).
\end{equation}
Here $\mathbf e_a$ is the $a$th coordinate vector in $\mathbb R^J$, with $a=1,2,3$ corresponding to arms $10,01,11$, and $\psi^{[p]}$ denotes the first $p$ coordinates of an infinite coefficient sequence. Assume that the propensity scores have the form
\[
 e_{a,n}(x)=\eta_a^{[P_n]\prime}x^{[P_n]},\quad a=1,\ldots,J,
 \qquad e_{0,n}(x)=1-\sum_{a=1}^J e_{a,n}(x),\quad J=3.
\]
The rows of $\eta$ are $\eta_a'$. The infinite sequences provide a common parameter space for the growing dimension of the nuisance parameters. To obtain sufficiently fast rates for the estimated nuisance parameters, I impose sparsity and use LASSO and post-LASSO to estimate the high-dimensional nuisance parameters. Let the first control variable be a constant ($X_1=1$) and the outcome intercept $\psi_1$ and the intercepts in the treatment regressions will be unpenalized and always retained in the post-LASSO step. Because of the required rate of convergence, the model class over which inference is uniform therefore restricts the sparsity of these coefficients. The following common sequence space also makes it possible to formulate Assumption \hyperref[A1]{1} as the number of controls grows. Fix $B_X\ge1$, $B_c>0$, and $a_k\downarrow0$, and set $\mathcal K=\{x\in[-B_X,B_X]^{\mathbb N}:x_1=1\}$ and
\[
 \mathcal C=\left\{b= \{b_j \}_{j \in \mathbb{N}} :\norm b_1\le B_c,\ \sum_{j>k}|b_j|\le a_k\text{ for all }k\ge1\right\}.
\]
 
 The common tail bound ensures that when we are dealing with only a finite number of the initial elements of the infinite-dimensional nuisance parameters (such as using $\psi^{[P_n]}$ instead of $\psi$) can bound how close their $\ell_1$ norms are (e.g., $||\psi||_1 \leq ||\psi^{[P_n]}||_1 + a_{P_n}$ ). Let $\omega=(\psi,\eta,\nu,F)$, where $F$ is the law of $X$ and $\nu$ lies in a fixed compact metric space $\mathcal H$ indexing the conditional error laws $\mathcal L_{\nu,n}(\varepsilon\mid D,X)$. The entire indexed family must satisfy the error conditions below. Set $\Gamma=\Theta\times\mathcal W$ where the nuisance space $\mathcal W$ consists of the tuples satisfying Condition~\ref{mt:condition}(i)--(iii) below. Thus $\gamma=(\theta,\omega)$ has the same interpretation as in Assumption \hyperref[A1]{1}, with $\Omega(\theta)=\mathcal W$. Here, $\Psi_n=\mathbb R^{P_n}$ and the finite-prefix orthogonalization matrix lies in $\mathbb R^{J\times P_n}$.

For the estimation formulas, I abbreviate $X_i^{[P_n]}$ by $X_i$. For any $b\in\mathbb R^{P_n}$ define $\|b\|_{1,-1}=\sum_{j=2}^{P_n}|b_j|$, excluding the intercept. Define the unpenalized criterion as
\begin{equation}\label{mt:ls}
 \hat Q(\theta,\psi)=\frac1{2n}\sum_{i=1}^n(Y_i-D_i'\theta-X_i'\psi)^2.
\end{equation}
Take $\lambda_\psi=C_\psi\sqrt{\log(P_n+n)/n}$ and $\lambda_\eta=C_\eta\sqrt{\log(P_n+n)/n}$ for sufficiently large fixed constants, with $C_\psi\ge2C_\varepsilon$ for the constant in Condition~\ref{mt:condition} below. Obtain a first step LASSO estimate using
\begin{align}
 (\hat\theta^{\,L},\hat\psi^{\,L})
 &=\argmin_{\theta\in\Theta,\,\psi\in\mathbb R^{P_n}}
       \hat Q(\theta,\psi)+\lambda_\psi\|\psi\|_{1,-1},\notag\\
 \hat\eta_a
 &=\argmin_{b\in\mathbb R^{P_n}}
       \frac1{2n}\sum_{i=1}^n(D_{ia}-X_i'b)^2+\lambda_\eta\|b\|_{1,-1}.
       \label{mt:node}
\end{align}
Let $\hat S=\operatorname{supp}(\hat\psi^{\,L})\cup\{1\}$, retaining the intercept regardless of selection, and perform the constrained post-LASSO refit
\begin{equation}\label{mt:postlasso}
 (\hat\theta^{\,PL},\hat\psi)
 =\argmin_{\theta\in\Theta,\,\psi\in\mathbb R^{P_n}:\,\psi_{\hat S^c}=0}
       \hat Q(\theta,\psi).
\end{equation}
Then we can use the post-LASSO coefficient $\hat\psi$ and the treatment-regression LASSO matrix $\hat\eta$, whose rows are $\hat\eta_a'$, in the orthogonal moments
\begin{align*}
 \hat M(\theta,\psi,\eta)
 =D_\theta\hat Q(\theta,\psi)-\eta D_\psi\hat Q(\theta,\psi)
 =-\frac1n\sum_{i=1}^n(D_i-\eta X_i)(Y_i-D_i'\theta-X_i'\psi).
\end{align*}
The initial constrained estimator $\hat{\theta}$ for Proposition~\ref{P5} is then the approximate minimizer satisfying
\begin{equation}\label{mt:gmm}
 \norm{\hat M(\hat\theta,\hat\psi,\hat\eta)}_2^2
 \le\inf_{\theta\in\Theta}\norm{\hat M(\theta,\hat\psi,\hat\eta)}_2^2+ o_p(1/n),
\end{equation}
since the $\hat\theta^{\,L}$ and $\hat\theta^{\,PL}$ are simply auxiliary estimates used to obtain $\hat\psi$.

\begin{condition}\label{mt:condition}
For fixed $c_X>0$, $0<\underline\pi<1/(J+1)$, and finite constants $C_\varepsilon,C_2>0$, the following hold.
\begin{enumerate}[label=(\roman*)]
\item $\psi,\eta_1,\ldots,\eta_J\in\mathcal C$, $\nu\in\mathcal H$, and $F$ is supported on $\mathcal K$, with
\begin{equation*}
 \mathbb E_F[(h'X^{[p]})^2]\ge c_X\norm h_2^2
 \quad(p\ge1,\ h\in\mathbb R^p).
\end{equation*}
\item $e_{a,n}(x)\ge\underline\pi$ for every $n$, $x\in\mathcal K$, and $a=0,\ldots,J$.
\item For non-stochastic bounds $s_{\psi,n},s_{\eta,n}$, $\norm{\psi^{[P_n]}}_0\le s_{\psi,n}$, $\sum_{a=1}^J\norm{\eta_a^{[P_n]}}_0\le s_{\eta,n},$ for every $n$, and
\begin{equation*}
 (J+s_{\psi,n}+s_{\eta,n})\log(P_n+n)=o(\sqrt n).
\end{equation*}
\item Observations are i.i.d. under \eqref{mt:model}, $X_i\sim F$, and
\[
 \mathbb E_{\gamma}[\varepsilon_i\mid D_i,X_i]=0,\qquad
 \sup_{n,\gamma\in\Gamma}\mathbb E_{\gamma}\varepsilon_i^2\le C_2.
\]
Along each $\{\gamma_n\}_{n \in \mathbb{N}} \in \Gamma(\gamma^*)$, the outcome LASSO selects $|\hat S|=O_p(J+s_{\psi,n})$ and equation~\eqref{mt:gmm} holds.
\item For each $\{\gamma_n\}_{n \in \mathbb{N}} \in \Gamma(\gamma^*)$, write $Z_i=(D_i',X_i^{[P_n]\prime})'$ and $V_{i,n}=D_i-\eta_nX_i^{[P_n]}$, where $\eta_n$ is the true finite-prefix matrix. Then
\begin{align}
 &\Pr_{\gamma_n}\left(\norm{\frac1n\sum_{i=1}^nZ_i\varepsilon_i}_\infty
       \le C_\varepsilon\sqrt{\log(P_n+n)/n}\right)\rightarrow1,\notag\\
 &\frac1{\sqrt n}\sum_{i=1}^n V_{i,n}\varepsilon_i
       \Rightarrow N(0,V_M(\gamma^*)),\label{mt:errorclt}\\
 &\frac1n\sum_{i=1}^n V_{i,n}V_{i,n}'\varepsilon_i^2
       \rightarrow_p V_M(\gamma^*),\label{mt:errorlln}
\end{align}
for a continuous symmetric matrix function $V_M$ on $\Gamma$ with
$\inf_{\gamma\in\Gamma}\lambda_{\min}(V_M(\gamma))>0$.
\end{enumerate}
\end{condition}


As in other settings with double-robust estimators, the orthogonalizing values of $\eta$ are the true propensity score coefficients. As a result, limiting values of the derivatives of the orthogonal moment conditions with respect to $\theta$ are based on the limiting covariance matrix of the error terms in these propensity scores as more and more covariates are included. Therefore, let $ M_{\theta}(\gamma^*)=\lim_{n \rightarrow \infty} \mathbb{E}_{\gamma_n}[V_{i,n}D_i']= \lim_{n \rightarrow \infty} \mathbb{E}_{\gamma_n}[V_{i,n}V_{i,n}']$ under $\{\gamma_n \}_{n \in \mathbb{N}} \in \Gamma(\gamma^*)$.

\begin{proposition}\label{mt:proposition}
Under Condition~\ref{mt:condition}, Assumptions 1 and 3 hold for the unpenalized criterion \eqref{mt:ls}, post-LASSO outcome estimate \eqref{mt:postlasso} and LASSO treatment estimates \eqref{mt:node}, and constrained GMM estimator \eqref{mt:gmm}. Consequently, Proposition~\ref{P5} applies and
\[
 \sqrt n(\widetilde\theta-\theta_n)
 \Rightarrow N(0,\Sigma(\gamma^*)).
\]
where $\Sigma(\gamma^*)=M_\theta(\gamma^*)^{-1}V_M(\gamma^*)M_\theta(\gamma^*)^{-1}$. Furthermore, $\hat\Sigma=\hat M_\theta^{-1}\hat V_M(\hat M_\theta^{-1})' \overset{p}{\rightarrow} \Sigma(\gamma^*)$, where
\[
 \hat M_\theta=\frac1n\sum_{i=1}^n(D_i-\hat\eta X_i)D_i',\qquad
 \hat V_M=\frac1n\sum_{i=1}^n\hat v_i\hat v_i'\hat u_i^2,\quad
 \hat v_i=D_i-\hat\eta X_i,\quad
 \hat u_i=Y_i-D_i'\hat\theta-X_i'\hat\psi.
\]
\end{proposition}

In part (i) of Condition~\ref{mt:condition}, the common tail bound makes the coefficient space compact and ensures that the outcome and treatment probabilities vary continuously as the coefficients change. It requires more than sparsity because coefficients on controls appearing later in the chosen ordering must have a collectively small contribution. The lower bound on $\mathbb E_F[(h'X^{[p]})^2]$ rules out arbitrarily small variation in linear combinations of controls with unit-length coefficient vectors, which helps control estimation error. Part (ii) requires every treatment arm, including the untreated arm, to remain possible at every control value. This preserves treatment variation after conditioning on the controls and identifies the treatment effects. Part (iii) limits how quickly the number of relevant controls can grow relative to the sample size. Together with the other conditions, its rate restriction makes the remaining errors from estimating the nuisance coefficients negligible for the limiting distribution of the treatment-effect estimator. Part (iv) imposes independent sampling and a conditionally mean-zero outcome error. Its bound on the number of selected controls ensures that the post-LASSO refit remains sufficiently small, without requiring selection of exactly the true set of relevant controls. Finally, part (v) links the error distribution to estimation and inference. Its first requirement bounds the sample correlations between regressors and errors at the scale of the LASSO penalty. Here, LASSO was used for the propensity score, and post-LASSO was used for the outcome model, but modifying the conditions to hold for a version where both use LASSO, post-LASSO, or another similar $ \ell_1$-penalized estimator such as the Dantzig selector would be straightforward. Post-LASSO offers the benefit of reducing the finite-sample bias in the estimated nuisance parameters, but it is also the reason why the bound on the number of controls selected by the outcome LASSO is needed in Part (iv).

\section{Asymptotic Power}\label{s: Power}

\cite{KETZ2018} analyzes the pointwise-in-$\beta$ power properties of their version of the CLR test when $\Theta$ is a product space and the subvector being tested $\beta$ is a scalar. They show that when the CLR test is pointwise similar, it is also admissible and extended Weighted Average Power (WAP) maximizing in the class of all tests with correct asymptotic size.\footnote{Extended-WAP-similar is defined \cite{montiel_olea_2020} to be a similar test that, for any tolerance $\epsilon > 0$, there exist some weighting of alternatives such that the test is within $\epsilon$ of being WAP-maximizing in the set of similar tests.} The core of their argument relies on showing that the test is similar with convex acceptance regions and using an argument based on the complete class result of \cite{MatthesTruax1967} to show that this means the CLR is admissible. This can then be connected to being WAP-maximizing using the results of \cite{montiel_olea_2020}, who shows that a test is similar and admissible if and only if it is similar and essentially WAP-maximizing (i.e., the limit of a sequence of WAP-maximizing tests). Here I discuss the challenges in establishing general conditions under which tests will be admissible and give examples of how both the Wald and CLR tests can be inadmissible.

\par 
Due to the product space condition, \cite{KETZ2018} is able to treat testing for a boundary point as a single special case (because the only boundary case is $H_0: \beta = 0$).\footnote{More specifically, for the boundary case, they must add a condition that $\alpha \leq 1/2$ and there is positive covariance of $\beta$ with the elements of $\delta$ which are also constrained to be nonnegative.}
To see why it is necessary to handle boundary points separately, first note that when testing points on the boundary, admissibility is not guaranteed by convex acceptance regions and similarity alone. This can be straightforwardly shown using the Wald test. Consider the class of tests $\mathcal{C}$ that asymptotically have size less than or equal to $\alpha$ under a data generating processes that induce $\sqrt{n}\hat{\Sigma}^{-1/2}(\Tilde{\theta} - \theta_0) \Rightarrow  N(0_J,I_J)$. Based on \cite{Muller2011}, for any test that is a function of this quasi-unconstrained estimator $\phi(\sqrt{n}\hat{\Sigma}^{-1/2}(\Tilde{\theta} - \theta_0))$, its WAP converges to the WAP of $\phi$ in the limiting problem. That is, its asymptotic WAP can be found by considering $\phi(\mathcal{Z})$ where $\mathcal{Z} \sim N(0_J,I_J)$. As a result of this, it is possible to show whether a test of this form is asymptotically WAP-maximizing in the class $\mathcal{C}$ by showing whether it is WAP-maximizing in this class for the limiting problem. For the Wald test, it is easy to find counterexamples where it is not WAP-maximizing in the limiting problem when the null value is on the boundary. Consider the following example: Let $\Theta = [0,\infty) \times (-\infty,\infty)$, $H_0: \; \beta = 0$, and the true value of $\beta$ is some $\beta_0 \geq 0$. The power of the Wald test in the limiting problem is $1- P(|\tilde{\beta}| \leq \sqrt{\Sigma_{\beta\beta}} z_{1-\alpha/2})$ where $\Tilde{\beta} \sim N(\beta_0,\Sigma_{\beta \beta})$ and $z_{1-\alpha/2}$ denotes the quantile from a standard normal distribution. If instead a right-sided test was used, then the power would be $1- P(\Tilde{\beta} \leq \sqrt{\Sigma_{\beta \beta}} z_{1-\alpha})$. Then the difference in the right-sided and the two-sided power is given by 
$P( \sqrt{\Sigma_{\beta \beta}} z_{1-\alpha} \leq \Tilde{\beta} \leq  \sqrt{\Sigma_{\beta \beta}} z_{1-\alpha/2}) - P(\Tilde{\beta} \leq -\sqrt{\Sigma_{\beta \beta}} z_{1-\alpha/2} ),$ which is positive for all $\beta_0 > 0$ and equal to zero when $\beta_0 = 0$. Therefore, the Wald test is inadmissible among the class $\mathcal{C}$. As the Wald test illustrates, a similar test with convex acceptance regions can still be dominated when the tested value lies on the boundary of the parameter of interest because a one-sided competitor exploits information about $\Theta$ when determining the rejection region that a two-sided statistic ignores.

\par 
A second complication in boundary cases is that, due to discontinuities in the asymptotic distribution of the test statistic, the test may not be similar. In the case of \cite{KETZ2018}, they are able to handle the special case of testing $\beta = 0$ because the only discontinuity occurs at probability one-half, so as long as $\alpha < 1/2$ this does not present an issue and the $\epsilon$-adjustment to the critical value is not needed. In fact, they note that for the example above with testing $H_0: \beta = 0$ when $\Theta = [0,\infty) \times (-\infty,\infty)$, the unadjusted CLR is identical to the one-sided test, meaning that the unadjusted CLR test dominates the Wald test when $\alpha < 1/2$. However, for a general parameter space, no such simple characterization of the discontinuities is available, so ensuring the test is similar becomes correspondingly more difficult. 

I now show how the CLR test can be inadmissible for all $\alpha \in (0,1)$ when $\Theta$ is not a product space in an example where $\beta$ and $\delta$ are scalars. Consider the limit experiment
$(\Tilde\beta,\Tilde\delta)'\sim N((\beta,\delta)',I_2)$ so $X = \Tilde\delta$ with
$\Theta=\{(\beta,\delta):|\beta|\le\delta\}$, and test $\beta=0$.
When $X<0$, minimizing under the null sets $d=0$, giving the
quadratic criterion $\Tilde\beta^2+X^2$. For any unrestricted
$b$, the closest feasible nuisance value is $d=|b|$.
Choosing the sign of $b$ to match $\Tilde\beta$ therefore gives
\[
 CLR(0,\Tilde\beta,I_2,X)
 =\Tilde\beta^2+X^2
  -\inf_{d\ge0}\bigl\{(|\Tilde\beta|-d)^2+(X-d)^2\bigr\}.
\]
The derivative of the expression being minimized is
$4d-2(|\Tilde\beta|+X)$, so its minimizer is
$d=\max\{(|\Tilde\beta|+X)/2,0\}$. Substituting gives
\[
 CLR(0,\Tilde\beta,I_2,X)
 =\begin{cases}
  0,&|\Tilde\beta|\le-X,\\
  (|\Tilde\beta|+X)^2/2,&|\Tilde\beta|>-X,
 \end{cases}
 \qquad X<0.
\]
Thus the statistic is exactly zero whenever $|\tilde{\beta}| < -X = - \tilde{\delta}$. Since $\tilde{\beta} \sim N(0,1)$ under the null and is independent of $X$, the probability of the statistic being equal to zero conditional on $X$ is $P(\chi_1^2 \le X^2)$. The conditional critical value is therefore zero precisely when
$X<0$ and $X^2\ge cv_{1-\alpha}(\chi_1^2)$, or equivalently
$X\le- z_{1-\alpha/2} $.
When the inequality is strict, rejecting only above zero has
conditional rejection probability below $\alpha$. Such values of
$X$ have positive probability under every value
$\delta\ge0$, for every $\alpha\in(0,1)$.
For $X\ge0$, the same minimization gives $\Tilde\beta^2$ when
$|\Tilde\beta|\le X$ and
$\Tilde\beta^2-(|\Tilde\beta|-X)^2/2$ otherwise, which is strictly
increasing in $|\Tilde\beta|$. Together with the formula for $X<0$,
this shows that the unadjusted CLR rejects if and only if $|\Tilde\beta|>  \max\bigl\{z_{1-\alpha/2},-X\bigr\},$ whereas the Wald test rejects whenever $|\Tilde\beta|>z_{1-\alpha/2}$.
Both tests control size at level $\alpha$ in this limiting experiment, but the CLR rejection
region is a strict subset of the Wald rejection region.
The difference, where
$|\Tilde\beta|>z_{1-\alpha/2}$ and
$X\le-|\Tilde\beta|$, has positive probability under every feasible
alternative because the Gaussian density is positive everywhere.
Thus the Wald test strictly dominates the CLR in pointwise
asymptotic power in this limit experiment, and adding the fixed
$\epsilon>0$ further restricts the CLR rejection region.  If $\beta$ were also constrained to be nonnegative, then the two-sided Wald would no longer dominate the CLR. However, an analogous argument could be made to show that the one-sided version of the Wald test would dominate the CLR when $\Theta = \{\theta: 0 \leq \beta \leq \delta \}$ when $0 < \alpha < 1/2$. In this case, the unadjusted CLR rejects when $\tilde{\beta} > \max \{ z_{1-\alpha},0, -X \}$ whereas the one-sided test rejects when $\tilde{\beta} > \max \{ z_{1-\alpha},0 \}$, which happens with strictly higher probability when the true value of $\beta$ is greater than zero. Therefore, this one-sided test strictly dominates the CLR when $0 < \alpha < 1/2$. In Appendix B of the Supplementary Materials, I show that a similar argument can also be made when studying the local power in monotone treatment application of Subsection~\ref{sub: MontoneTreatment} where $\Theta=\{\theta\in[0,B]^3 : \theta_{10}\leq\theta_{11},\ \theta_{01}\leq\theta_{11}\}.$ I show that the one-sided test will weakly dominate the CLR for testing whether the separate interventions had any effect, and I compare whether there is any practical difference in power in the simulations in Subsection~\ref{s: Simulations}.

\section{Empirical Application and Simulations}\label{s: Application}

I illustrate the proposed inference methods using data from \cite{BJORKMAN_GUARISO2027}, who evaluate the effectiveness of two educational interventions implemented by Pratham in Assam, India. The first intervention, "Learning Camps" (LC), provides targeted instruction during school hours, grouping students according to their learning levels rather than their grades. The second intervention, "Study Groups" (SG), organizes community-led learning activities outside school hours to encourage additional practice and peer learning. The authors conduct a randomized controlled trial involving 200 villages, randomly assigned to four equally sized groups: Learning Camps only, Study Groups only, both interventions, and a control group. The study follows a sample of 5,726 primary school children, with baseline data collected in 2018 and endline data collected approximately 17 months later. I focus on their mathematics achievement, measured using standardized scores from the Annual Status of Education Report (ASER) assessments. As in Subsection~\ref{sub: MontoneTreatment}, treatment effects can be estimated for the effect of Learning Camps alone \(\theta_{LC}\), the effect of Study Groups alone \(\theta_{SG}\), and the effect of implementing both interventions \(\theta_{LC\&SG}\). The authors find that neither intervention separately has a statistically significant effect on learning outcomes, whereas the combined effect is statistically significant when some baseline controls are included. To match the setting in Subsection~\ref{sub: MontoneTreatment}, I aggregate the data at the village level. Table~\ref{tab:aser_village_estimates_math} contains the estimates using the village-level data, including no controls, including a larger set of 62 control variables with no penalty on the control coefficients, and using post-LASSO with the same 62 control variables.\footnote{All variables measured at baseline by \cite{BJORKMAN_GUARISO2027} are included: baseline test scores (ASER, Test A and IRT scores in math and language, plus indicators for ASER levels), child demographics (age and grade indicators and gender), schooling and study habits (presence in school at baseline, studying outside school, study groups, who the child studies with, and liking school), and interactions of the six baseline test scores with grade and gender. This yields 64 variables, two of which (the top ASER math and reading level indicators) are dropped because they are exact linear combinations of the others.} This gives similar estimates as in the primary specification of \cite{BJORKMAN_GUARISO2027}, in which 6 baseline controls\footnote{This includes only the baseline value of the outcome, age, gender and grade indicators.} are included. They estimate that the combined effect is an increase of approximately 0.11 standard deviations in math test scores. 


\begin{table}[htbp]
\centering
\caption{Estimated Effects on ASER Math Scores}
\label{tab:aser_village_estimates_math}
\scalebox{.8}{
\begin{threeparttable}
\begin{tabular}{lcccccc}
\toprule
 & \multicolumn{3}{c}{Quasi-unconstrained, $\tilde{\theta}$} & \multicolumn{3}{c}{Constrained, $\hat{\theta}$} \\
\cmidrule(lr){2-4} \cmidrule(lr){5-7}
 & LC & SG & LC \& SG & LC & SG & LC \& SG \\
\midrule
No controls & $-$0.001 (0.079) & 0.000 (0.089) & 0.141 (0.081) & 0.000 & 0.001 & 0.141 \\
62 controls, unpenalized & 0.040 (0.043) & $-$0.001 (0.048) & 0.104 (0.040) & 0.041 & 0.000 & 0.105 \\
62 controls, post-LASSO & 0.057 (0.034) & 0.028 (0.037) & 0.125 (0.033) & 0.057 & 0.028 & 0.125 \\
\bottomrule
\end{tabular}
\begin{tablenotes}[flushleft]
\footnotesize
\item \textit{Notes:} Effects on village-mean ASER math scores, in control-group standard deviations. $\tilde{\theta}$ is the quasi-unconstrained estimate, with its HC2 standard error in parentheses, and $\hat{\theta}$ is the constrained estimate. For the LASSO step: $\lambda = \mathrm{SD}(\mathrm{dep.\ var.})\sqrt{\log(P_n+n)/n}$.
\end{tablenotes}
\end{threeparttable}
}
\end{table}

This application provides a useful setting for evaluating whether more informative  inference results can be obtained if we are willing to impose restrictions on the relationships between the treatment effects. In particular, we can impose the monotonicity restrictions \(0\leq\theta_{LC}\leq\theta_{LC\&SG}\) and \(0\leq\theta_{SG}\leq\theta_{LC\&SG}\), reflecting the assumption that each intervention weakly improves learning outcomes and that their combination is at least as effective as either intervention alone. These restrictions generate a non-product parameter space, with potentially binding inequalities given the small estimated effects of the individual interventions.

In addition to testing whether each of the interventions each had a statistically significant effect, it is interesting to test whether the combined effect equals the effect of Learning Camps alone (\(\theta_{LC\&SG}=\theta_{LC}\)) or Study Groups alone (\(\theta_{LC\& SG}=\theta_{SG}\)). Both hypotheses place the parameter on the boundary of the parameter space, since they correspond to binding monotonicity restrictions. I also test the null hypothesis of additive treatment effects, \(\theta_{LC\&SG}=\theta_{LC}+\theta_{SG}\), under which the effect of implementing both interventions equals the sum of their individual effects. Rejection in the direction \(\theta_{LC\&SG}>\theta_{LC}+\theta_{SG}\) would indicate complementarity, whereas rejection in the opposite direction would indicate substitutability. Unlike the other hypotheses, additivity is consistent with parameters both on the boundary and in the interior of the monotonicity-constrained parameter space, depending on whether the individual treatment effects are zero or positive. The subvector tests of Section~\ref{s: InferenceForSubvectors} test hypotheses about the leading elements of the parameter vector. To test a linear combination $r'\theta$, we can reparameterize as $\phi = R\theta$, where $R$ is an invertible $3 \times 3$ matrix whose first row is $r'$, so that $\beta = \phi_1 = r'\theta$ is the parameter of interest and the other two elements of $\phi$ are the nuisance parameters $\delta$. The estimator becomes $\tilde{\phi} = R\tilde{\theta}$ with asymptotic variance $R\Sigma R'$, and the parameter space becomes $\Phi = \{R\theta : \theta \in \Theta\}$  with the CLR statistics and critical values computed in the $\phi$ coordinates exactly as for a test of the first element of the parameter vector. For example, to test additivity, take $\phi = (\tau, \theta_{LC}, \theta_{SG})'$ with $\tau = \theta_{LC\&SG} - \theta_{LC} - \theta_{SG}$. The constraints become $\theta_{LC}, \theta_{SG} \geq 0$ and $\tau \geq -\min(\theta_{LC}, \theta_{SG})$.

\begin{table}[!ht]
\centering
\caption{P-values for the Effects on Math Scores}
\label{tab:aser_village_pvalues}
\scalebox{.8}{
\begin{threeparttable}
\begin{tabular}{lcccccc}
\toprule
 & LC & SG & LC \& SG & \shortstack{LC \& SG\\$-$ SG} & \shortstack{LC \& SG\\$-$ LC} & Additivity \\
\midrule
\multicolumn{7}{l}{\textit{Panel A: No controls}} \\
QUE Wald & 0.994 & 0.996 & 0.081 & 0.123 & 0.083 & 0.242 \\
QUE Right-sided & 0.503 & 0.498 & 0.040 & 0.061 & 0.041 &  \\
Wald with $\hat{\theta}$ & 1.000 & 0.991 & 0.080 & 0.123 & 0.083 & 0.244 \\
QUE CLR & 1.000 & 0.498 & 0.040 & 0.061 & 0.041 & 0.237 \\
Fan and Shi QLR & 1.000 & 0.634 & 0.073 & 0.093 & 0.071 & 0.262 \\
Li (2025) & 1.000 & 1.000 & 0.243 & 0.409 & 0.403 & 0.717 \\
\addlinespace
\multicolumn{7}{l}{\textit{Panel B: 62 controls}} \\
QUE Wald & 0.093 & 0.447 & $<$0.001 & 0.006 & 0.034 & 0.403 \\
QUE Right-sided & 0.047 & 0.223 & $<$0.001 & 0.003 & 0.017 &  \\
Wald with $\hat{\theta}$ & 0.093 & 0.447 & $<$0.001 & 0.006 & 0.034 & 0.403 \\
QUE CLR & 0.047 & 0.223 & $<$0.001 & 0.003 & 0.017 & 0.404 \\
Fan and Shi QLR & 0.208 & 1.000 & 0.011 & 0.015 & 0.076 & 0.319 \\
Li (2025) & 0.740 & 1.000 & 0.057 & 0.131 & 0.463 & 0.766 \\
\bottomrule
\end{tabular}
\begin{tablenotes}[flushleft]
\footnotesize
\item \textit{Notes:} Sample and estimates are as in Table \ref{tab:aser_village_estimates_math}. In Panel B, the QUE tests use post-LASSO estimates: the controls selected by either LASSO step are refit by least squares, with HC2 leverage from the selected controls. \cite{FAN2023} and \cite{LI2025} use all 62 controls without penalization, but they are profiled out for \cite{LI2025}. The Wald test with $\hat{\theta}$ uses the constrained estimate with the same variance as the test with $\tilde{\theta}$. CLR p-values are the share of the simulated conditional null distribution at or above the statistic minus $\epsilon = 10^{-6}$, from 1,000 quadrature points. \cite{FAN2023}'s QLR test uses 2,000 low-discrepancy draws and gives decisions at a fixed level, so its p-value is the smallest $\alpha$ at which it rejects, with $\tau = 0.9\alpha$. Both the QUE tests and the QLR use the HC2 covariance estimator. \cite{LI2025}'s p-values are the share of 1,000 leverage-adjusted bootstrap draws at or above the smallest statistic among points in the constraint set that satisfy the null.
\end{tablenotes}
\end{threeparttable}
}
\end{table}

Comparing the proposed CLR, one-sided, and Wald tests, which all use $\tilde{\theta}$, allows us to assess whether exploiting the monotonicity restrictions yields greater power in practice. For all results with the CLR, $\epsilon = 10^{-6}$ is used. In the simulations below, this addition to the critical values does not affect rejection in any of the draws. In fact, increasing $\epsilon$ does not impact rejection until it is between $10^{-2}$ and $10^{-3}$. Alongside the QUE tests, I report the traditional Wald test, which uses the constrained estimate $\hat{\theta}$, and two recent alternative inference methods that can also accommodate the constraints in this setting. \cite{FAN2023} develop tests of linear restrictions on a parameter vector subject to linear inequality constraints, so the parameter space for $\theta$ fits with their formal results. Their Wald statistic is built from the constrained estimator, and their Quasi-Likelihood Ratio (QLR) statistic compares the fit of the objective with and without the null imposed. The null distributions of both depend on which inequalities bind, which is not known, so their critical values take the most conservative value over a confidence set for the parameters that determine this, with a Bonferroni-type correction. Their score test instead projects the score onto the set of values the tested parameter can take. I focus on including their QLR test as it is most similar to the CLR used here. \cite{LI2025} proposes the proximal bootstrap, which forms a confidence set for the full parameter vector by comparing a local change in the objective with a bootstrap critical value from an unconstrained quadratic problem. When the true parameters satisfy the constraints, the constraints affect this procedure only through the point at which the bootstrap is evaluated, and intervals for a single effect are obtained by projecting the joint set. 


Looking at the p-values in Table~\ref{tab:aser_village_pvalues}, we can see that the results for the tests of additivity are consistently not statistically significant, but for tests that the combined effect is different from SG or LC alone, which test is used and whether baseline controls are included does influence whether the difference is statistically significant at standard thresholds. I don't include the one-sided test for the additivity null since it isn't sign-restricted, but for the other hypotheses we can see that it gives the same p-value as the CLR in all but one case. When testing $H_0: \theta_{LC} = 0$ with no controls, the CLR gives p-values of 1, which is greater than the p-values of the Wald and one-sided tests. This relates to the previously mentioned points about the potential for point masses in the asymptotic distribution of the CLR statistic, as it is the one case where the CLR statistic is exactly zero. Besides this, the CLR or one-sided tests generally give the lowest p-values except for testing additivity with many controls. On the other hand, the proximal bootstrap gives the largest p-values. Using $\hat{\theta}$ instead of $\tilde{\theta}$ makes either no difference or only a slight difference as the QUE is either inside $\Theta$ or only slightly outside $\Theta$.

\subsection{Simulations}\label{s: Simulations}


In order to better understand the size and power properties of these tests in this setting, I conduct simulations calibrated to the data from \cite{BJORKMAN_GUARISO2027}, focusing again on the ASER math test scores. They are calibrated to the village-level data, so each simulated sample has 200 villages, with every control held fixed at its observed village mean. The outcome is generated from an OLS regression of the math scores on the treatment indicators and the six controls that \cite{BJORKMAN_GUARISO2027} originally included in their main results. Simulated outcomes are then calculated by resampling the residuals from this regression with replacement. Since the results in Subsection~\ref{sub: MontoneTreatment} allow for non-random treatment assignment, I consider cases with both random treatment assignment and treatment assignment that is only exogenous conditional on the controls. More specifically, for village $i$ with controls $X_i$, the indicators $D_{LC,i}$ and $D_{SG,i}$ for being offered each treatment are drawn independently of each other, and of the outcome residual, with $\Pr(D_{LC,i} = 1 \mid X_i) = p_{LC}(X_i)$ and $\Pr(D_{SG,i} = 1 \mid X_i) = p_{SG}(X_i)$. In the equal probabilities specification, $p_{LC}(X_i) = p_{SG}(X_i) = 1/2$ for every village, so assignment doesn't depend on the controls at all. In the take-up assignment specification, $p_{k}(X_i) = \Lambda(a_k + X_i'\gamma_k)$ for $k \in \{LC, SG\}$, where $\Lambda$ is the logistic function. The slopes $\gamma_k$ are estimated by a logit LASSO of child-level take-up on the control variables at the child level. The probability of being offered each treatment ranges from about 0.42 to 0.59 across villages. Of course, in real life, the variables that determine child-level selection into a program and village-level selection into a program would not necessarily be the same, so this is not intended to be a method for determining how the program would actually roll out in a setting without random assignment. Rather, it serves as a simple, data-driven way to estimate propensity scores that differs from estimating $\hat{\eta}$ because it uses disaggregated data and the logit functional form.

\begin{table}[htbp]
\centering
\caption{Size Results}
\label{tab:sim_size_math}
\resizebox{\textwidth}{!}{%
\begin{threeparttable}
\begin{tabular}{lcccccccccccc}
\toprule
 & \multicolumn{6}{c}{Equal probabilities} & \multicolumn{6}{c}{Take-up assignment} \\
\cmidrule(lr){2-7} \cmidrule(lr){8-13}
 & \multicolumn{3}{c}{6 controls} & \multicolumn{3}{c}{62 controls} & \multicolumn{3}{c}{6 controls{}} & \multicolumn{3}{c}{62 controls{}} \\
\cmidrule(lr){2-4} \cmidrule(lr){5-7} \cmidrule(lr){8-10} \cmidrule(lr){11-13}
 & Far & Close & Boundary & Far & Close & Boundary & Far & Close & Boundary & Far & Close & Boundary \\
\midrule
\multicolumn{13}{l}{\textit{Panel A: SG}} \\
QUE Wald & 5.4 & 5.3 & 5.1 & 5.6 & 5.6 & 5.3 & 4.8 & 4.7 & 4.7 & 5.0 & 5.0 & 4.7 \\
QUE Right-sided  & 5.4 & 5.4 & 5.4 & 5.5 & 5.5 & 5.5 & 5.0 & 5.0 & 5.0 & 5.5 & 5.6 & 5.6 \\
Wald with $\hat{\theta}$ & 5.4 & 2.9 & 2.9 & 5.6 & 3.2 & 3.2 & 4.8 & 2.5 & 2.5 & 5.0 & 2.7 & 2.7 \\
QUE CLR & 5.4 & 5.4 & 5.4 & 5.6 & 5.6 & 5.5 & 4.8 & 4.9 & 5.0 & 5.0 & 5.2 & 5.6 \\
Fan and Shi QLR & 4.6 & 5.2 & 5.0 & 4.8 & 5.1 & 5.2 & 4.6 & 4.7 & 4.6 & 4.4 & 4.9 & 5.0 \\
Li (2025) & 0.6 & 0.7 & 1.5 & 0.6 & 0.7 & 1.3 & 0.6 & 0.5 & 1.1 & 0.6 & 0.6 & 1.1 \\
\addlinespace
\multicolumn{13}{l}{\textit{Panel B: LC \& SG $-$ LC}} \\
QUE Wald & 5.6 & 5.5 & 5.2 & 5.9 & 5.7 & 5.3 & 5.4 & 5.4 & 5.2 & 5.4 & 5.4 & 5.1 \\
QUE Right-sided & 5.2 & 5.2 & 5.2 & 5.7 & 5.7 & 5.6 & 5.3 & 5.3 & 5.3 & 5.6 & 5.6 & 5.5 \\
Wald with $\hat{\theta}$ & 5.6 & 2.7 & 2.7 & 5.9 & 2.9 & 2.8 & 5.4 & 2.5 & 2.4 & 5.4 & 2.6 & 2.6 \\
QUE CLR & 5.6 & 5.5 & 5.2 & 5.9 & 5.5 & 5.6 & 5.4 & 5.4 & 5.3 & 5.4 & 5.7 & 5.5 \\
Fan and Shi QLR & 5.2 & 5.0 & 4.7 & 5.1 & 4.1 & 4.2 & 5.0 & 4.8 & 4.4 & 4.9 & 4.1 & 4.1 \\
Li (2025) & 0.6 & 0.7 & 1.3 & 0.7 & 0.8 & 1.2 & 0.6 & 0.6 & 1.1 & 0.7 & 0.8 & 1.3 \\
\addlinespace
\multicolumn{13}{l}{\textit{Panel C: LC \& SG $-$ SG $-$ LC}} \\
QUE Wald & 5.2 & 5.1 & 5.0 & 5.3 & 5.1 & 4.9 & 4.4 & 4.4 & 4.2 & 4.9 & 4.9 & 4.6 \\
Wald with $\hat{\theta}$ & 5.2 & 2.0 & 0.5 & 5.3 & 2.1 & 0.5 & 4.4 & 2.0 & 0.4 & 4.9 & 1.9 & 0.6 \\
QUE CLR & 5.2 & 5.0 & 4.7 & 5.3 & 5.1 & 4.5 & 4.4 & 4.6 & 4.2 & 4.9 & 4.9 & 4.6 \\
Fan and Shi QLR & 4.7 & 3.4 & 1.2 & 5.1 & 3.3 & 1.0 & 4.3 & 2.9 & 1.2 & 4.4 & 2.9 & 1.2 \\
Li (2025) & 0.7 & 0.7 & 1.4 & 0.5 & 0.6 & 1.2 & 0.6 & 0.5 & 1.0 & 0.6 & 0.6 & 1.1 \\
\bottomrule
\end{tabular}
\begin{tablenotes}[flushleft]
\footnotesize
\item \textit{Notes:} Rejection rates, in percent, of 5\% tests of true null hypotheses, from 5,000 replications per design. True effects are additive, with ($\theta_{LC}, \theta_{SG}, \theta_{LC\&SG}$) equal to (0.22, 0.22, 0.44) in Far, (0.22, 0.04, 0.26) in Close and (0.22, 0, 0.22) in Boundary. All tests are conducted as in Table~\ref{tab:aser_village_pvalues}.
\end{tablenotes}
\end{threeparttable}
}
\end{table}

I also consider both the case where only the six controls that appear in the outcome model are used and the case where the large set of 62 controls are included. The true treatment effects are chosen to consider cases where the tests are far from, close to, and at the boundary. For the size results, $\theta_{LC} = 0.22$ and the effects are additive, $\theta_{LC\&SG} = \theta_{LC} + \theta_{SG}$, with $\theta_{SG}$ equal to 0.22 (far from the boundary), 0.04 (about one standard error from it) or 0. Each null hypothesis is tested at its true value. The results in Table~\ref{tab:sim_size_math} show that all of the tests generally succeed in controlling size, with each of them over-rejecting by at most one percentage point across the specifications. For the post-LASSO estimator under take-up assignment, the refitted control coefficients do have notably lower bias than the first-step estimates, where they are shrunk towards zero, indicating the standard LASSO would likely have performed worse than post-LASSO.

\begin{table}[!ht]
\centering
\caption{Power Results}
\label{tab:sim_power_math}
\scalebox{.8}{
\begin{threeparttable}
\begin{tabular}{lcccccccccccc}
\toprule
 & \multicolumn{6}{c}{Equal probabilities} & \multicolumn{6}{c}{Take-up assignment} \\
\cmidrule(lr){2-7} \cmidrule(lr){8-13}
 & \multicolumn{3}{c}{6 controls} & \multicolumn{3}{c}{62 controls} & \multicolumn{3}{c}{6 controls{}} & \multicolumn{3}{c}{62 controls{}} \\
\cmidrule(lr){2-4} \cmidrule(lr){5-7} \cmidrule(lr){8-10} \cmidrule(lr){11-13}
True LC \& SG: & 0.09 & 0.14 & 0.19 & 0.09 & 0.14 & 0.19 & 0.09 & 0.14 & 0.19 & 0.09 & 0.14 & 0.19 \\
\midrule
\multicolumn{13}{l}{\textit{Panel A: LC \& SG $-$ SG $= 0$}} \\
QUE Wald & 21.7 & 65.7 & 94.5 & 22.0 & 65.7 & 94.1 & 20.9 & 65.6 & 94.6 & 22.1 & 65.8 & 94.2 \\
QUE Right-sided & 32.3 & 76.6 & 97.2 & 32.5 & 76.1 & 97.0 & 31.7 & 76.6 & 97.2 & 32.3 & 76.6 & 97.3 \\
Wald with $\hat{\theta}$ & 21.2 & 65.5 & 94.5 & 21.5 & 65.7 & 94.1 & 20.5 & 65.6 & 94.6 & 21.6 & 65.9 & 94.2 \\
QUE CLR & 32.3 & 76.6 & 97.2 & 32.5 & 76.1 & 96.9 & 31.7 & 76.6 & 97.2 & 32.3 & 76.6 & 97.3 \\
Fan and Shi QLR & 24.9 & 71.6 & 96.5 & 19.1 & 56.6 & 87.8 & 24.9 & 71.9 & 96.8 & 18.6 & 56.9 & 88.8 \\
Li (2025) & 7.7 & 38.7 & 81.7 & 5.6 & 24.5 & 62.1 & 7.2 & 37.4 & 82.0 & 5.1 & 25.2 & 62.0 \\
\addlinespace
\multicolumn{13}{l}{\textit{Panel B: LC \& SG $-$ LC $= 0$}} \\
QUE Wald & 56.2 & 91.0 & 99.3 & 56.1 & 90.6 & 99.3 & 55.5 & 90.8 & 99.5 & 55.8 & 90.6 & 99.4 \\
QUE Right-sided & 68.5 & 95.2 & 99.7 & 68.1 & 94.8 & 99.7 & 67.3 & 95.1 & 99.8 & 67.6 & 94.8 & 99.7 \\
Wald with $\hat{\theta}$ & 55.1 & 90.8 & 99.2 & 55.0 & 90.4 & 99.3 & 54.3 & 90.7 & 99.5 & 54.5 & 90.5 & 99.3 \\
QUE CLR & 68.5 & 95.2 & 99.7 & 68.1 & 94.8 & 99.7 & 67.3 & 95.1 & 99.8 & 67.6 & 94.8 & 99.7 \\
Fan and Shi QLR & 63.4 & 93.8 & 99.7 & 48.4 & 84.0 & 97.6 & 62.9 & 93.9 & 99.7 & 48.7 & 83.6 & 97.8 \\
Li (2025) & 30.1 & 72.7 & 96.3 & 20.1 & 52.8 & 85.3 & 30.2 & 72.6 & 96.6 & 20.1 & 53.4 & 85.6 \\
\addlinespace
\multicolumn{13}{l}{\textit{Panel C: LC \& SG $-$ SG $-$ LC $= 0$}} \\
QUE Wald & 13.0 & 38.7 & 70.7 & 12.8 & 39.0 & 69.5 & 12.8 & 38.7 & 71.2 & 13.0 & 38.6 & 70.5 \\
Wald with $\hat{\theta}$ & 2.8 & 23.6 & 63.9 & 2.8 & 23.9 & 62.9 & 2.6 & 23.7 & 64.4 & 2.8 & 23.9 & 63.8 \\
QUE CLR & 13.1 & 39.1 & 70.7 & 12.9 & 39.3 & 69.6 & 12.9 & 38.8 & 71.3 & 13.5 & 39.0 & 70.6 \\
Fan and Shi QLR & 7.0 & 33.9 & 70.4 & 5.0 & 22.2 & 52.9 & 7.0 & 33.7 & 70.6 & 5.2 & 22.7 & 52.4 \\
Li (2025) & 2.5 & 13.4 & 40.5 & 2.0 & 8.7 & 25.0 & 2.4 & 13.4 & 40.4 & 1.5 & 8.4 & 25.3 \\
\bottomrule
\end{tabular}
\begin{tablenotes}[flushleft]
\footnotesize
\item \textit{Notes:} Rejection rates, in percent, of 5\% tests of the null hypothesis in each panel, from 5,000 replications per design. Column headings give the true LC \& SG effect, with $\theta_{LC} = 0$ and $\theta_{SG} = 0.04$. All other settings are as in Table \ref{tab:sim_size_math}.
\end{tablenotes}
\end{threeparttable}
}
\end{table}

We can see that the Wald test with the constrained estimator becomes conservative when the treatment effect subvector is close to the boundary. In general, the presence of constraints may lead tests that use the constrained estimate to either over- or under-reject, depending on the context. In this case, the results are qualitatively similar to the simplest case of a one-dimensional parameter constrained to lie on an interval discussed in Section \ref{s: Power}, leading to the traditional Wald test being conservative. The QLR test of \cite{FAN2023} is somewhat conservative for the test of additivity, likely due to the Bonferroni correction, and \cite{LI2025} is conservative across the specifications, likely due to projecting the joint confidence set for testing the linear restrictions.\footnote{It is worth noting that I make several choices in implementing the proximal bootstrap, most of which make this version less conservative than other versions the paper would allow. First, the controls are profiled out. Applied directly, Remark 3 of \cite{LI2025} projects over every coefficient in the objective, which would be 10 with the six controls and 66 with 62 controls, so such a test would almost never reject. Second, I use Theorem 1 rather than Theorem 2, which is reasonable because the true $\theta$ lies in $\Theta$ and is also the unconstrained minimizer. Third, I leave the local deviations $h$ unrestricted ($\kappa = \infty$), which \cite{LI2025}'s Remark 2 notes is the least conservative choice. Fourth, we intersect the projected set with $\Theta$, which lets the test reject near the boundary more often. Finally, I rescale each residual using the HC2 adjustment before resampling the scores. This is a finite-sample correction the paper doesn't discuss, but it is used to match the HC2 covariance used by the other tests.}


Table~\ref{tab:sim_power_math} contains power results for the same tests. Here, $\theta_{LC} = 0$ is on its boundary and $\theta_{SG} = 0.04$, while $\theta_{LC\&SG}$ rises from 0.09 to 0.19, so that the true value of each of the three differences being tested is farther from zero by an increasing amount. The Wald test using the constrained estimator generally has lower power than the Wald test using the QUE, likely because $\theta$ is on the boundary at $\theta_{LC} = 0$. The QLR test has slightly lower but similar power when only 6 controls are included, but significantly lower power when all 62 controls are included. The proximal bootstrap generally has the lowest power of the tests. Interestingly, when comparing the CLR and the one-sided test, the tests either have identical rejection rates, or the CLR is very slightly more conservative. In the Supplementary Materials, I show that a similar argument made in Section~\ref{s: Power} can be made to prove that the one-sided test dominates the CLR in terms of local asymptotic power in the monotone treatment application. However, while there are cases where dominance is strict, additional simulations in Appendix B align with those presented here, showing that the difference in power is very small. When comparing the Wald and CLR tests for additivity, the advantage of using the CLR is also very small. On the one hand, these results suggest that the cost to the CLR from the potential point mass in its distribution is small. On the other hand, they also suggest that, unlike sign constraints, which have a clear impact on power, the benefit of the CLR accounting for interactive constraints is often much smaller.

\section{Discussion}\label{s: Discussion}

An important limitation of the method is the rate of convergence conditions required on the initial constrained estimate. When Neyman orthogonality is used, $\sqrt{n}$-consistency is not required for the entire vector, but the nuisance-parameter estimator must satisfy rate conditions tied to the sensitivity of the derivatives of the criterion function in order for the conditions in Assumption \hyperref[A3]{3.2} to hold. This typically requires imposing structure on the nuisance parameter, for example, sparsity of the high-dimensional vector of controls. Additionally, these convergence rate conditions can fail when proximity to the boundary coincides with weak identification. The application in Section~\ref{s: Application} is a case where constraints are imposed for efficiency rather than identification. In some cases, constraints may be imposed for identification and the constrained estimate may still be $\sqrt{n}$-consistent (e.g., imposing sign constraints when a scalar coefficient's magnitude is identified), but if a boundary point of $\Theta$ corresponds to a point where $\theta$ becomes unidentified, then the constrained estimator will generally not be $ \sqrt{n}$-consistent under sequences approaching the boundary at a sufficiently fast rate. In this case, a method like \cite{LI2025} that can accommodate constraints and weak identification would be preferred. Given the lower power of the proximal bootstrap relative to the tests studied here, looking for more powerful tests for subvector inference in the presence of constraints and weak identification, and for tests that can handle weak identification and high-dimensionality, are interesting directions for future work. 
\par
Important theoretical questions remain relating to the power of these tests, including characterizing the conditions on the parameter space under which the CLR test can be (at least approximately) similar and admissible. While the simulation results generally find the CLR to have superior power to Wald test, this is because it is generally indistinguishable from the one-sided test in this setting. Furthermore, as illustrated in Section~\ref{s: Power}, it is possible for the CLR to be dominated by the Wald or by the one-sided test due to the potential for the CLR to be conservative. This suggests that the CLR benefits from increased power by taking into account bounds on the parameter of interest, but the case for using constraints involving the nuisance parameters in the test statistic seems much less clear. Related to this, orthogonalization and conditioning offer two distinct approaches to handling nuisance parameters, and as noted in Remark~\ref{r: interactionHighDim}, if there are constraints involving the nuisance parameter and the parameter of interest, then orthogonalizing with respect to the nuisance parameter is potentially ignoring useful information about what values of the parameter of interest are feasible. However, the fact that these interactive constraints can also lead the CLR to be conservative means that further work is needed to understand how and when to make use of constraints involving nuisance parameters to improve the power of tests.

\bigskip
\noindent
\textbf{Acknowledgments}: I am grateful for helpful comments and feedback from Adam McCloskey, Carlos Martins-Filho, Philipp Ketz, and Xiaodong Liu. GPT-6 Astra was used for text editing and for development and verification of author-written proofs. Claude Opus 5.5 was used for coding assistance in the simulation study and empirical application, but its output was augmented, edited, and verified. Application data is publicly available via \href{https://data.mendeley.com/datasets/mtf2x29p69/1}{https://data.mendeley.com/datasets/mtf2x29p69/1}.

{\footnotesize
\bibliographystyle{econ-aea}
\bibliography{references}}


\section*{Appendix A. Proofs of the Main Results}\phantomsection\label{ApA}

\textbf{Notational Notes}: Convergence of sequences of sets in meant in the Painlev´e-Kuratowski sense, unless otherwise noted.

\noindent\textbf{Proof of Proposition~\ref{grc:verification}} 

\begin{journalappendixmath}
\begin{proof}
Fix an arbitrary sequence $\{\gamma_G\}_{G\in\mathbb N}=\{(\theta_G,F_G)\}_{G\in\mathbb N}\in\Gamma(\gamma^*).$ All probabilities, expectations, and stochastic orders below refer to the $G$th experiment under $\gamma_G$. Write $\theta_G=(\beta_{X,G}',\xi_G')'$, $\xi_G=(\tau_G,u_G,v_G,w_G)'$, and let $\Omega_{b,G}$ be its random-coefficient covariance matrix. Put $Q_G(\theta)=\mathbb{E}_{\gamma_G}q(W;\theta)$. Constants in the bounds below can be chosen independently of $G$ and of this sequence.
The set $\Theta$ is compact and convex. For a weakly convergent sequence of regressors, expectations of the bounded continuous functions $X'X$ and $\left\lVert aI_m+ZHZ'\right\rVert_F^2$ converge. Their inequalities in Condition~\ref{grc:primitive}(i) are therefore preserved in the limit, for each fixed $(a,H)$. Thus $\mathcal F$ is closed in that compact space, and $\Gamma=\Theta\times\mathcal F$ is compact. If $\gamma_G\to\gamma^*$, the cluster laws also converge weakly. Indeed, represent $Y$ given $(X,Z)$ as
$X\beta_{X,G}+C_Z(\theta_G)^{1/2}U$, where $U\sim N(0,I_m)$ is independent of $(X,Z)$ and the square root is symmetric positive definite.

The map from $(X,Z,U,\theta)$ to $(Y,X,Z)$ is continuous, and the law of $(X,Z,U)$ converges weakly because $F_G\to F^*$ and the independent law of $U$ is fixed. Independence across clusters gives the corresponding convergence for any fixed number of clusters. This verifies Assumption~\ref{A1}. Compactness of the regressor support gives a constant $K_Z$ such that $\left\lVert Z\right\rVert\le K_Z$. Consequently,
$\underline\tau I_m\preceq C_Z(\theta)\preceq C_+I_m, \qquad C_+=\overline\tau+BK_Z^2,\qquad \theta\in\Theta.$
Write a parameter perturbation as
$h=(h_X',h_\tau,h_u,h_v,h_w)'.$
Let $H_h$ be the symmetric matrix with diagonal entries $h_u,h_v$ and off-diagonal entry $h_w$. Then
$\left\lVert C_Z(\theta+h)-C_Z(\theta)\right\rVert \le |h_\tau|+K_Z^2\left\lVert H_h\right\rVert\le K_0\left\lVert h\right\rVert$
for a finite constant $K_0$. Choose $\rho>0$ with $K_0\rho<\underline\tau/2$, and let $\mathcal N=\Theta+\{h:\left\lVert h\right\rVert\le\rho\}$ be the resulting compact convex neighborhood. Every $C_Z(\theta)$ on $\mathcal N$ has smallest eigenvalue at least $\underline\tau/2$.

For each covariance coordinate $\xi_k$, the derivative matrix $A_k=\partial C_Z/\partial\xi_k$ is bounded uniformly over the regressor support. The identities
$\frac{\partial C_Z^{-1}}{\partial\xi_k}=-C_Z^{-1}A_kC_Z^{-1}$ and$ \frac{\partial\log\det C_Z}{\partial\xi_k} =\operatorname{tr}(C_Z^{-1}A_k)$
show that derivatives of $q$ through order three are sums of products of bounded matrices and at most two factors of $Y-X\beta_X$. Here $\operatorname{tr}$ denotes the trace. Since $X$ and candidate $\beta_X$ are bounded on the specified sets, there is a finite $K$ such that
\begin{equation}\label{grc:envelope}
 \sup_{\theta\in\mathcal N}\left\lVert D^jq(W;\theta)\right\rVert
 \le M(W):=K(1+\left\lVert Y\right\rVert^2),\qquad j=0,1,2,3.
\end{equation}
For higher-order derivative arrays use the norm induced by Euclidean norms. The conditional Gaussian representation above and bounded means and covariances give
$\sup_G\mathbb{E}_{\gamma_G}\left\lVert Y\right\rVert^r<\infty \text{for every fixed }r\ge1, \sup_G\mathbb{E}_{\gamma_G}M(W)^r<\infty.$

I give the argument for each $f(W;\theta)=D^jq(W;\theta)$, $j=0,1,2$. The mean-value formula and \eqref{grc:envelope} imply
$\left\lVert f(W;\theta)-f(W;\vartheta)\right\rVert \le M(W)\left\lVert \theta-\vartheta\right\rVert, \theta,\vartheta\in\Theta.$
Here $\vartheta$ denotes a second candidate point solely for this comparison. For $\delta>0$, compactness supplies a finite grid $\theta^{(1)},\ldots,\theta^{(N_\delta)}$ such that every point in $\Theta$ lies within $\delta$ of a grid point. Define the centered sample average
$T_G^f(\theta)=\frac1G\sum_{g=1}^G f(W_g;\theta) -\mathbb{E}_{\gamma_G}f(W;\theta), \overline M_G=\frac1G\sum_{g=1}^G M(W_g).$
Independence of the $G$ clusters in each experiment, the uniform second-moment bound, and Chebyshev's inequality give a finite constant $K_f$ such that
\[
 \Pr_{\gamma_G}\!\left(\max_{k\le N_\delta}
       \left\lVert T_G^f(\theta^{(k)})\right\rVert>\epsilon\right)
 \le\sum_{k=1}^{N_\delta}
       \frac{\mathbb{E}_{\gamma_G}\left\lVert T_G^f(\theta^{(k)})\right\rVert^2}{\epsilon^2}
 \le\frac{N_\delta K_f}{G\epsilon^2}.
\]
For matrix or array entries the same bound follows by using the Frobenius norm, which dominates the operator norm. Interpolation gives
\[
\sup_{\theta\in\Theta}\left\lVert T_G^f(\theta)\right\rVert \le\max_{k\le N_\delta}\left\lVert T_G^f(\theta^{(k)})\right\rVert +\delta\bigl(\overline M_G+\mathbb{E}_{\gamma_G}M(W)\bigr).
\]
Writing $K_M=\sup_G\mathbb{E}_{\gamma_G}M(W)$ and applying Markov's inequality to the second term yields
\[
\limsup_{G\to\infty}\Pr_{\gamma_G}\!\left( \sup_{\theta\in\Theta}\left\lVert T_G^f(\theta)\right\rVert>\epsilon\right) \le\frac{4\delta K_M}{\epsilon}.
\]
Letting $\delta\downarrow0$ establishes
\begin{equation}\label{grc:ulln}
 \sup_{\theta\in\Theta}\left\lVert D^j\hat Q(\theta)-D^jQ_G(\theta)\right\rVert=o_p(1),
 \qquad j=0,1,2.
\end{equation}
Differentiation under the expectation is justified by \eqref{grc:envelope}. For $f(W;\theta)=s(W;\theta)s(W;\theta)'$, the identity
$aa'-bb'=(a-b)a'+b(a-b)'$ bounds its norm by $M(W)^2$ and its Lipschitz envelope by $2M(W)^2$. Their second moments are uniformly bounded by the Gaussian eighth-moment bound. The same grid argument therefore proves the uniform law for this outer product as well.

Index $\xi$ by $k=0,1,2,3$, and define
\[
 A_0=I_m, A_k=ZE_kZ' \text{ for }k=1,2,3,
 E_1=\begin{pmatrix}1&0\\0&0\end{pmatrix},
 E_2=\begin{pmatrix}0&0\\0&1\end{pmatrix},
 E_3=\begin{pmatrix}0&1\\1&0\end{pmatrix}.
\]
For a generic cluster in the $G$th experiment, let $e_G=Y-X\beta_{X,G}$ be the residual after subtracting the true conditional mean and let $C_G=C_Z(\theta_G)$ be its conditional covariance. The components of $s_G=s(W;\theta_G)$ are
$s_{G,\beta_X}=-X'C_G^{-1}e_G, s_{G,\xi_k}=\frac12\left( \operatorname{tr}(C_G^{-1}A_k) -e_G'C_G^{-1}A_kC_G^{-1}e_G\right).$
Conditionally on $(X,Z)$, $e_G\sim N(0,C_G)$. For symmetric matrices $R,S$, Gaussian moments give $ \mathbb{E}_{\gamma_G}[e_G'Re_G\mid X,Z]=\operatorname{tr}(RC_G)$, $ \operatorname{Cov}_{\gamma_G}(e_G'Re_G,e_G'Se_G\mid X,Z)=2\operatorname{tr}(RC_GSC_G)$, and $ \mathbb{E}_{\gamma_G}[e_G(e_G'Re_G)\mid X,Z]=0.$
Substitution gives $\mathbb{E}_{\gamma_G}[s_G\mid X,Z]=0$. The odd Gaussian moments make the cross blocks between the mean and covariance parameters zero. Averaging over $(X,Z)$ gives
\begin{equation*}
 \mathbb{E}_{\gamma_G}s_Gs_G'=\mathcal I(\gamma_G)
 =\operatorname{diag}\!\left(
 \mathbb{E}_{F_G}[X'C_G^{-1}X],
 \left[\frac12\mathbb{E}_{F_G}\operatorname{tr}
       (C_G^{-1}A_kC_G^{-1}A_\ell)\right]_{k,\ell=0}^3\right).
\end{equation*}
Here $\operatorname{diag}$ forms a block diagonal matrix. Direct differentiation also gives the same expected Hessian. In particular, its mean block is $X'C_G^{-1}X$, its cross block has entries linear in $e_G$ and hence conditional expectation zero, and its covariance block is
\begin{align*}
 \left.\frac{\partial^2q}{\partial\xi_k\partial\xi_\ell}\right|_{\theta_G}
 ={}&-\frac12\operatorname{tr}(C_G^{-1}A_\ell C_G^{-1}A_k)
 &+\frac12 e_G'\bigl(C_G^{-1}A_\ell C_G^{-1}A_kC_G^{-1}
                 +C_G^{-1}A_kC_G^{-1}A_\ell C_G^{-1}\bigr)e_G.
\end{align*}
Taking conditional expectations reduces this expression to
$\frac12\operatorname{tr}(C_G^{-1}A_kC_G^{-1}A_\ell)$. Thus
$D^2Q_G(\theta_G)=\mathcal I(\gamma_G)$.

Let $\mathcal I_{\beta_X\beta_X}$ and $\mathcal I_{\xi\xi}$ denote the information blocks for $\beta_X$ and $\xi$, respectively. For any $h_X\in\mathbb R^p$,
$h_X'\mathcal I_{\beta_X\beta_X}(\gamma_G)h_X \ge C_+^{-1}\mathbb{E}_{F_G}\left\lVert Xh_X\right\rVert^2 \ge(c_X/C_+)\left\lVert h_X\right\rVert^2.$
For $h=(a,h_1,h_2,h_3)'$, put $H=\sum_{k=1}^3h_kE_k$ and $L_Z=aI_m+ZHZ'$. The covariance block obeys $ h'\mathcal I_{\xi\xi}(\gamma_G)h = $
$$ \frac12\mathbb{E}_{F_G}\left\lVert C_G^{-1/2}L_ZC_G^{-1/2}\right\rVert_F^2
 \ge\frac{1}{2C_+^2}\mathbb{E}_{F_G}\left\lVert L_Z\right\rVert_F^2
 \ge\frac{c_C}{2C_+^2}(a^2+\left\lVert H\right\rVert_F^2)
 \ge\frac{c_C}{2C_+^2}\left\lVert h\right\rVert^2.$$
The inverse square root is the symmetric one. The first inequality follows from
\[
\left\lVert L_Z\right\rVert_F=\left\lVert C_G^{1/2}(C_G^{-1/2}L_ZC_G^{-1/2})C_G^{1/2}\right\rVert_F \le C_+\left\lVert C_G^{-1/2}L_ZC_G^{-1/2}\right\rVert_F.
\]
Hence $\mathcal I(\gamma_G)\succeq\kappa I_d$ for
$\kappa=\min(c_X/C_+,c_C/(2C_+^2))>0$.
The same formulas and bounded regressors give a uniform upper bound. Their integrands are jointly continuous in the regressor pair and the parameter on a compact set. Weak convergence $F_G\to F^*$ and $\theta_G\to\theta^*$ consequently imply
$\mathcal I(\gamma_G)\rightarrow\mathcal I(\gamma^*).$
For example, each entry has the form $\int f(X,Z,\theta_G)\,dF_G(X,Z)$, and its difference from the limit is bounded by
\[
\sup_{(X,Z)\in\mathcal K}|f(X,Z,\theta_G)-f(X,Z,\theta^*)| +
\]
\[
\sup_{(X,Z)\in\mathcal K}\left|\int f(X,Z,\theta^*)\,dF_G(X,Z)-\int f(X,Z,\theta^*)\,dF^*(X,Z)\right|\to0.
\]

For a candidate $\theta=(\beta_X',\xi')'$, write $C=C_Z(\theta)$. The conditional expectation of the criterion difference gives $ Q_G(\theta)-Q_G(\theta_G)=$
\begin{align*}
\frac12\mathbb{E}_{F_G}\bigl[&\operatorname{tr}(C^{-1}C_G)-m
                  -\log\det(C^{-1}C_G)
                  +(X(\beta_X-\beta_{X,G}))'C^{-1}X(\beta_X-\beta_{X,G})\bigr].
\end{align*}
The eigenvalues $\lambda_j$ of the symmetric matrix $C^{-1/2}C_GC^{-1/2}$ lie in
$[\underline\tau/C_+,C_+/\underline\tau]$. The continuous ratio
$(\lambda-1-\log\lambda)/(\lambda-1)^2$, extended at $\lambda=1$ by $1/2$, has a positive minimum $c_0$ on that interval. Therefore
$$\operatorname{tr}(C^{-1}C_G)-m-\log\det(C^{-1}C_G)
 =\sum_{j=1}^m(\lambda_j-1-\log\lambda_j)$$
$$ \ge c_0\left\lVert C^{-1/2}(C_G-C)C^{-1/2}\right\rVert_F^2
 \ge\frac{c_0}{C_+^2}\left\lVert C_G-C\right\rVert_F^2.$$
Apply Condition~\ref{grc:primitive}(i) with $a=\tau-\tau_G$ and $H=\Omega_b-\Omega_{b,G}$. Together with the mean bound this yields
\begin{equation*}
 Q_G(\theta)-Q_G(\theta_G)\ge c_Q\left\lVert \theta-\theta_G\right\rVert^2,
 \qquad c_Q=\frac12\min\left(\frac{c_X}{C_+},\frac{c_0c_C}{C_+^2}\right)>0.
\end{equation*}
Here $\left\lVert \Omega_b-\Omega_{b,G}\right\rVert_F^2=(u-u_G)^2+(v-v_G)^2+2(w-w_G)^2$ dominates the squared Euclidean distance between the three covariance entries. Put
$a_G=\sup_{\theta\in\Theta}|\hat Q(\theta)-Q_G(\theta)|=o_p(1)$.
Feasibility of $\theta_G$ and approximate minimization give
$c_Q\left\lVert \hat\theta-\theta_G\right\rVert^2 \le Q_G(\hat\theta)-Q_G(\theta_G) \le2a_G+r_G=o_p(1).$
Thus $\hat\theta-\theta_G=o_p(1)$.

Let $\Delta_G=\hat\theta-\theta_G$, $S_G=D\hat Q(\theta_G)$, and $L_G=G^{-1}\sum_{g=1}^G M(W_g)=O_p(1)$. The third-derivative bound gives
$$\sup_{0\le v\le1}\left\lVert D^2\hat Q(\theta_G+v\Delta_G) -D^2\hat Q(\theta_G)\right\rVert \le L_G\left\lVert \Delta_G\right\rVert=o_p(1).$$
Also, \eqref{grc:ulln} and the information identity imply
$D^2\hat Q(\theta_G)=\mathcal I(\gamma_G)+o_p(1)$.
Thus every Hessian on this segment is at least $(\kappa/2)I_d$ with probability tending to one. The integral Taylor formula then gives
$$ r_G\ge\hat Q(\hat\theta)-\hat Q(\theta_G) =S_G'\Delta_G+\int_0^1(1-v)\Delta_G'D^2\hat Q(\theta_G+v\Delta_G)\Delta_G\,dv$$
$$\ge-\left\lVert S_G\right\rVert\left\lVert \Delta_G\right\rVert+(\kappa/4)\left\lVert \Delta_G\right\rVert^2.$$
The zero mean and independence of the cluster scores give
$\mathbb{E}_{\gamma_G}\left\lVert S_G\right\rVert^2 =G^{-1}\operatorname{tr}\mathcal I(\gamma_G)=O(G^{-1}).$
Consequently $\left\lVert S_G\right\rVert=O_p(G^{-1/2})$. Solving the preceding quadratic inequality gives, on the event of positive local curvature,
$\left\lVert \Delta_G\right\rVert\le\frac{4}{\kappa}\left\lVert S_G\right\rVert +2\sqrt{r_G/\kappa}=O_p(G^{-1/2}),$
since $Gr_G=o_p(1)$. This proves Assumption~\hyperref[A2]{2.1}.

For any fixed nonzero $\ell\in\mathbb R^d$, let
$\sigma_G^2=\ell'\mathcal I(\gamma_G)\ell\ge\kappa\left\lVert \ell\right\rVert^2$.
The third absolute moment of $\ell's(W;\theta_G)$ is uniformly bounded by \eqref{grc:envelope}. The Lyapunov ratio for the row of $G$ independent scores is therefore
\[
 \frac{\sum_{g=1}^G\mathbb{E}_{\gamma_G}|\ell's(W_g;\theta_G)|^3}
      {(G\sigma_G^2)^{3/2}}=O(G^{-1/2})\to0.
\]
The triangular-array Lyapunov theorem gives the scalar normal limit for every such $\ell$. By the Cram\'er--Wold device and information convergence,
$\sqrt G\,D\hat Q(\theta_G)\Rightarrow N(0,\mathcal I(\gamma^*)).$
For the Hessian, the uniform law yields directly
$\left\lVert D^2\hat Q(\theta_G)-\mathcal I(\gamma^*)\right\rVert \le\sup_{\theta\in\Theta}\left\lVert D^2\hat Q(\theta)-D^2Q_G(\theta)\right\rVert +\left\lVert \mathcal I(\gamma_G)-\mathcal I(\gamma^*)\right\rVert\to_p0.$
These are Assumptions~\hyperref[A2]{2.2}--\hyperref[A2]{2.3}. Define $H_G=D^2\hat Q(\theta_G)$ and the gradient remainder
$R_G=D\hat Q(\hat\theta)-S_G-H_G\Delta_G$.
The integral formula for the gradient and the preceding Lipschitz bound give
\begin{align*}
 R_G&=\int_0^1\bigl(D^2\hat Q(\theta_G+v\Delta_G)-H_G\bigr)\Delta_G\,dv,\\
 \left\lVert R_G\right\rVert&\le\int_0^1vL_G\left\lVert \Delta_G\right\rVert^2\,dv
       =\tfrac12 L_G\left\lVert \Delta_G\right\rVert^2=O_p(G^{-1})=o_p(G^{-1/2}),\\
 \left\lVert D^2\hat Q(\hat\theta)-H_G\right\rVert
 &\le L_G\left\lVert \Delta_G\right\rVert=O_p(G^{-1/2})=o_p(1).
\end{align*}
With exact analytical derivatives, these verify both requirements of Assumption~\hyperref[A2]{2.4}.
Let $B_G(\theta)=\mathbb{E}_{\gamma_G}[s(W;\theta)s(W;\theta)']$.
The outer-product uniform law and its Lipschitz envelope imply
\begin{align*}
 \left\lVert \hat V(\hat\theta)-\mathcal I(\gamma^*)\right\rVert
 \le{} \sup_{\theta\in\Theta}\left\lVert \hat V(\theta)-B_G(\theta)\right\rVert +2\mathbb{E}_{\gamma_G}M(W)^2\left\lVert \Delta_G\right\rVert
   +\left\lVert \mathcal I(\gamma_G)-\mathcal I(\gamma^*)\right\rVert=o_p(1).
\end{align*}
Here $B_G(\theta_G)=\mathcal I(\gamma_G)$. This proves Assumption~\hyperref[A2]{2.5}.
Set $\hat H_G=D^2\hat Q(\hat\theta)$. The Hessian limit and local expansion give $\hat H_G\to_p\mathcal I(\gamma^*)$, so its inverse exists with probability tending to one and is $O_p(1)$. On this event, substituting the gradient expansion in the one-step definition gives
$$ \widetilde\theta-\theta_G
 =-\hat H_G^{-1}S_G
   +\hat H_G^{-1}(\hat H_G-H_G)\Delta_G
   -\hat H_G^{-1}R_G =-\hat H_G^{-1}S_G+o_p(G^{-1/2}).$$
The score central limit theorem and Slutsky's theorem establish the asserted normal limit. The sandwich estimator satisfies
$\hat\Sigma=\hat H_G^{-1}\hat V(\hat\theta)\hat H_G^{-1} \to_p\Sigma(\gamma^*)=\mathcal I(\gamma^*)^{-1}.$
Define the estimator and covariance arbitrarily on the vanishing-probability event where the Hessian is singular. Every conclusion above holds along the arbitrary convergent sequence fixed at the outset, including sequences with singular $\Omega_{b,G}$.
\end{proof}
\end{journalappendixmath}

\noindent\textbf{Proof of Proposition~\ref{P4}}
\begin{journalappendixmath}
\renewcommand{\qedsymbol}{$\blacksquare$}
\begin{proof}
Fix $\{\gamma_n\}_{n\in\mathbb N}
=\{(\beta_n,\delta_n,\omega_n)\}_{n\in\mathbb N}
\in\Gamma(\gamma^*)$.
Assumption~\ref{A2} implies
$\hat\Sigma\overset{p}{\rightarrow}\Sigma(\gamma^*)$, where
$\Sigma(\gamma^*)$ is positive definite. Thus $\hat\Sigma$ is positive
definite with probability approaching one. I establish the algebraic
identities below on this event; any fixed measurable convention on
its complement has no effect on the limiting probabilities.

Write $\hat C_n=\hat\Sigma_{\delta\beta}\hat\Sigma_{\beta\beta}^{-1}$
and let
$\hat\Lambda_n=\hat\Sigma_{\delta\delta}
-\hat\Sigma_{\delta\beta}\hat\Sigma_{\beta\beta}^{-1}
 \hat\Sigma_{\beta\delta}$ be the Schur complement.
Set $S_{D,n}=\sqrt n(D_{\beta_n}-\delta_n)$,
$\Tilde Z_n=\sqrt n(\Tilde\beta-\beta_n)$, and
$x_n=\sqrt n(X_n-\delta_n+\hat C_n\beta_n)$.
The inverse formula for a partitioned matrix gives, for
$z\in\mathbb R^K$ and $x,d\in\mathbb R^L$,
\[
 \begin{pmatrix}z\\x+\hat C_nz-d\end{pmatrix}'\hat\Sigma^{-1}
 \begin{pmatrix}z\\x+\hat C_nz-d\end{pmatrix}
 =z'\hat\Sigma_{\beta\beta}^{-1}z
    +(x-d)'\hat\Lambda_n^{-1}(x-d).
\]
Since
$\sqrt n(\Tilde\delta-\delta_n)=x_n+\hat C_n\Tilde Z_n$,
taking the infimum over $d\in S_{D,n}$ gives
\[
 CLM(\beta_n,\Tilde\beta,\hat\Sigma/n,X_n)
 =\Tilde Z_n'\hat\Sigma_{\beta\beta}^{-1}\Tilde Z_n
       +a_n(\beta_n),
 \qquad
 a_n(\beta_n)=\inf_{d\in S_{D,n}}
       (x_n-d)'\hat\Lambda_n^{-1}(x_n-d).
\]
This is the nuisance term appearing in the subsection's explanation.
It is finite because $0_L\in S_{D,n}$ and is fixed conditional on
$(X_n,\hat\Sigma)$.

Replacing $\Tilde\beta$ by
$n^{-1/2}\hat{\mathcal Z}_b+\beta_n$ in the conditional Gaussian
reference experiment leaves $x_n$ and $a_n(\beta_n)$ unchanged. Hence
\[
 CLM(\beta_n,n^{-1/2}\hat{\mathcal Z}_b+\beta_n,\hat\Sigma/n,X_n)
 =\hat{\mathcal Z}_b'\hat\Sigma_{\beta\beta}^{-1}
       \hat{\mathcal Z}_b+a_n(\beta_n).
\]
Conditional on $(X_n,\hat\Sigma)$,
$\hat{\mathcal Z}_b\sim N(0_K,\hat\Sigma_{\beta\beta})$, so the
quadratic term has exactly the $\chi_K^2$ distribution. Consequently,
$$cv_{1-\alpha}\!\left(
 CLM(\beta_n,n^{-1/2}\hat{\mathcal Z}_b+\beta_n,
            \hat\Sigma/n,X_n)\right)
 =cv_{1-\alpha}(\chi_K^2)+a_n(\beta_n).$$
The term $a_n(\beta_n)$ therefore cancels from the rejection rule:
\[
\begin{split}
 &\mathbf1\!\left\{
 CLM(\beta_n,\Tilde\beta,\hat\Sigma/n,X_n)>
 cv_{1-\alpha}\!\left(
 CLM(\beta_n,n^{-1/2}\hat{\mathcal Z}_b+\beta_n,
                   \hat\Sigma/n,X_n)\right)\right\}\\
 &\qquad=\mathbf1\!\left\{
 \Tilde Z_n'\hat\Sigma_{\beta\beta}^{-1}\Tilde Z_n
             >cv_{1-\alpha}(\chi_K^2)\right\}.
\end{split}
\]
Thus the CLM and subvector Wald tests give the same decision on the
positive-definite event, for every feasible null subvector.

Under Assumption~\ref{A2},
$(\Tilde Z_n,\hat\Sigma_{\beta\beta})
\Rightarrow 
(\mathcal Z_b,\Sigma_{\beta\beta}(\gamma^*))$, where
$\mathcal Z_b\sim N(0_K,\Sigma_{\beta\beta}(\gamma^*))$.
The continuous mapping theorem implies
$\Tilde Z_n'\hat\Sigma_{\beta\beta}^{-1}\Tilde Z_n
\Rightarrow \chi_K^2$.
Continuity at $cv_{1-\alpha}(\chi_K^2)$ gives
$CP_n(\gamma_n)\to1-\alpha$ for the Wald test. The equality of
decisions outside an event of vanishing probability gives the same
conclusion for the CLM test. These conclusions are uniform over $\Gamma$. Otherwise, for one of the tests, there would exist $\rho>0$, sample sizes $n_k\to\infty$, and $\gamma_{n_k}\in\Gamma$ with $|CP_{n_k}(\gamma_{n_k})-(1-\alpha)|\ge\rho$.
Compactness in Assumption~\ref{A1} supplies a further subsequence
converging to some $\gamma^*$. The preceding argument on that
subsequence is a contradiction. Thus, for each test,
$\sup_{\gamma\in\Gamma}|CP_n(\gamma)-(1-\alpha)|\to0$, implying
$AsySz=AsyMaxCP=1-\alpha$.
The same compactness argument and covariance consistency give
$\sup_{\gamma\in\Gamma}P_\gamma
(\lambda_{\min}(\hat\Sigma)\le0)\to0$.
The exact equality of decisions on the complementary event
therefore also establishes uniform asymptotic equivalence.
\end{proof}
\end{journalappendixmath}

\noindent\textbf{Proof of Proposition~\ref{P3}}
\begin{journalappendixmath}
\renewcommand{\qedsymbol}{$\blacksquare$}
\begin{proof}
Fix the positive adjustment $\epsilon$ in Proposition~\ref{P3}.
Let $\{\gamma_n\}_{n\in\mathbb N}
=\{(\beta_n,\delta_n,\omega_n)\}_{n\in\mathbb N}$ be an arbitrary
sequence in $\Gamma$, and take an arbitrary subsequence.
Set $S_n=\sqrt n(\Theta-\theta_n)$,
$S_{D,n}=\sqrt n(D_{\beta_n}-\delta_n)$, and
$N_n=\{0_K\}\times S_{D,n}$. The sets $S_n$ and $N_n$ are closed, contain zero, and satisfy $N_n\subseteq S_n$. We can construct a further subsequence along which both sets converge by considering their distance functions. For a nonempty closed set $A\subseteq\mathbb R^J$, let $d(x,A)=\inf_{a\in A}\lVert x-a\rVert_2$, where
$\lVert\cdot\rVert_2$ is the Euclidean norm. Locally uniform
convergence means uniform convergence on every bounded ball.

First, compactness in Assumption~\ref{A1} supplies a further
subsequence along which $\gamma_n\to\gamma^*$. All subsequent
selections are from this subsequence. For either $A_n=S_n$ or
$A_n=N_n$, zero is feasible, so
$0\le d(x,A_n)\le\lVert x\rVert_2$.
The triangle inequality gives
$d(x,A_n)\le\lVert x-y\rVert_2+d(y,A_n)$. Interchanging $x$ and $y$
then gives $|d(x,A_n)-d(y,A_n)|\le\lVert x-y\rVert_2.$ Thus the distance functions are uniformly bounded on each bounded ball, and their values at nearby arguments differ by at most the distance between those arguments. List the points with rational coordinates in $\mathbb R^J$ as $r_1,r_2,\ldots$. At $r_1$, the pairs $(d(r_1,S_n), d(r_1,N_n))$ form a bounded sequence in $\mathbb R^2$,
so they have a convergent subsequence. From that subsequence,
select a further subsequence on which the corresponding pairs at
$r_2$ converge, and continue. To obtain one subsequence, select
its $j$th index from the subsequence constructed at step $j$,
requiring each selected index to exceed the preceding one.
For each fixed $r_i$, all selected indices from step $i$ onward
then belong to the subsequence constructed at step $i$.
Consequently, both distance functions converge at every rational
point along this single subsequence. Denote its sample sizes by $m$.

This convergence is uniform on every bounded ball. To see this,
fix a ball $B_R=\{x\in\mathbb R^J\mid\lVert x\rVert_2\le R\}$ and
$h>0$. Choose finitely many rational points $r_1,\ldots,r_q$
(relabeling them for this argument) such that every point of
$B_R$ lies within distance $h$ of one of them. For either sequence
$A_m=S_m$ or $A_m=N_m$, the preceding distance inequality implies
\[
 \sup_{x\in B_R}|d(x,A_m)-d(x,A_\ell)|
 \le 2h+\max_{1\le i\le q}
             |d(r_i,A_m)-d(r_i,A_\ell)|.
\]
The maximum tends to zero as $m,\ell\to\infty$ because it involves
only finitely many convergent sequences. Since $h$ can be
arbitrarily small, the distance functions are uniformly Cauchy
on $B_R$ and hence converge uniformly there. This holds for every
$R$ and for both sequences of sets. Write their locally uniform
limits as $d_S$ and $d_N$.

I next show that each limiting function is itself the distance
function of a nonempty closed set. For either sequence $A_m$,
write $d_A$ for its limiting function and define
$A=\{a\in\mathbb R^J\mid d_A(a)=0\}$.
The limit satisfies $d_A(0_J)=0$ and
$|d_A(x)-d_A(y)|\le\lVert x-y\rVert_2$.
Thus $A$ contains zero and is closed. For every $a\in A$,
$d_A(x)\le\lVert x-a\rVert_2$, so $d_A(x)\le d(x,A)$.
For the reverse inequality, fix $x$ and choose a nearest point
$a_m\in A_m$. Such a point exists because a sequence approaching the
infimum defining $d(x,A_m)$ can be restricted to a bounded ball,
and a convergent subsequence has its limit in the closed set $A_m$.
Since zero is feasible, $ \lVert x-a_m\rVert_2=d(x,A_m)\le\lVert x\rVert_2$ and $\lVert a_m\rVert_2\le2\lVert x\rVert_2.$ The points $a_m$ therefore have a convergent subsequence, say
$a_{m_j}\to a$. Along this subsequence,
\[
 |d_A(a)-d(a_{m_j},A_{m_j})|
 \le |d_A(a)-d(a,A_{m_j})|+\lVert a-a_{m_j}\rVert_2
 \rightarrow0.
\]
Because $d(a_{m_j},A_{m_j})=0$, we have $a\in A$, and $ d(x,A)\le\lVert x-a\rVert_2
 =\lim_j d(x,A_{m_j})=d_A(x).$ Hence $d_A(x)=d(x,A)$. Applying this construction to both limits defines nonempty closed sets $S$ and $N$, each containing zero.

Finally, $N_m\subseteq S_m$ implies
$d(x,S_m)\le d(x,N_m)$ and hence $d(x,S)\le d(x,N)$.
If $x\in N$, this inequality gives $d(x,S)=0$, so $N\subseteq S$.
Moreover, for $x=(b,d)\in\mathbb R^K\times\mathbb R^L$,
$d(x,N_m)\ge\lVert b\rVert_2$ because every point in $N_m$ has
its first $K$ coordinates equal to zero. Passing to the limit
shows that every point of $N$ also has its first $K$ coordinates
equal to zero. Thus $N=\{0_K\}\times S_D$, where
$S_D=\{d\in\mathbb R^L\mid(0_K,d)\in N\}$ is closed and contains zero.
Since
$d((0_K,d),N_m)=d(d,S_{D,m})$ and
$d((0_K,d),N)=d(d,S_D)$, locally uniform convergence for $N_m$
also gives locally uniform convergence of the distance functions
of $S_{D,m}$.
We have therefore obtained a common further subsequence along which $ \{\gamma_m\}_{m\in\mathbb N}\in\Gamma(\gamma^*),$ $
 S_m\to S,$ and $ N_m\to N=\{0_K\}\times S_D.$ In particular, $S_{D,m}\to S_D$ and $\{0_K\}\times S_D\subseteq S$. Both limiting sets contain zero. It does not require identifying $S_D$ with an unscaled slice at $\beta^*$ or with the slice of $S$.

Write $\Sigma=\Sigma(\gamma^*)$,
$C=\Sigma_{\delta\beta}\Sigma_{\beta\beta}^{-1}$, and
$\Lambda=\Sigma_{\delta\delta}
-\Sigma_{\delta\beta}\Sigma_{\beta\beta}^{-1}\Sigma_{\beta\delta}$.
Set $\Tilde Z_m=\sqrt m(\Tilde\beta-\beta_m)$,
$\Tilde X_m=\sqrt m(\Tilde\delta-\delta_m)$, and
$\hat C_m=\hat\Sigma_{\delta\beta}\hat\Sigma_{\beta\beta}^{-1}$.
Define $x_m$ by the first equality below. The definition of $X_m$
gives the second equality, $x_m=\sqrt m(X_m-\delta_m+\hat C_m\beta_m)
       =\Tilde X_m-\hat C_m\Tilde Z_m.$ Assumption~\ref{A2} and Slutsky's theorem give
\[
 (\Tilde Z_m,x_m,\hat\Sigma)
 \Rightarrow (\mathcal Z_b,\mathcal X,\Sigma),
 \qquad
 \mathcal X=\mathcal Z_d-C\mathcal Z_b,\qquad
 \begin{pmatrix}\mathcal Z_b\\\mathcal Z_d\end{pmatrix}
       \sim N(0_J,\Sigma).
\]
Here $\mathcal Z_b\in\mathbb R^K$ and
$\mathcal Z_d,\mathcal X\in\mathbb R^L$.
Since
$\operatorname{Cov}(\mathcal Z_b,\mathcal X)
=\Sigma_{\beta\delta}-\Sigma_{\beta\beta}C'=0$ and
$\operatorname{Var}(\mathcal X)=\Lambda$,
$\mathcal Z_b$ and $\mathcal X$ are independent.
These are joint, unconditional convergence statements.
As in the proof of Proposition~\ref{P4}, covariance consistency
makes the probability of a positive-definite $\hat\Sigma$
approach one. Any fixed measurable convention on that event does not
affect the limiting probabilities.

For $x\in\mathbb R^L$, $z\in\mathbb R^K$, and a symmetric positive
definite $J\times J$ matrix $Y$, define
\[
\begin{split}
 f_m(x,Y,z)
 ={}\inf_{d\in S_{D,m}}
       \begin{pmatrix}z\\x-d\end{pmatrix}'Y
       \begin{pmatrix}z\\x-d\end{pmatrix}
 -\inf_{(b,d)\in S_m}
       \begin{pmatrix}z-b\\x-d\end{pmatrix}'Y
       \begin{pmatrix}z-b\\x-d\end{pmatrix}.
\end{split}
\]
Let $f(x,Y,z)$ be the same expression with $S_{D,m}$ and $S_m$
replaced by $S_D$ and $S$.
The feasible-set inclusion implies $f_m\ge0$ and $f\ge0$.
These functions are finite because zero is feasible.
For deterministic $(u_m,Y_m,z_m)\to(u,Y,z)$ with $Y$ positive
definite, Lemma~\hyperref[LA1]{A1} gives convergence of both
infima, and hence
\begin{equation}\label{clr:sequential-map}
 f_m(u_m,Y_m,z_m)\rightarrow f(u,Y,z).
\end{equation}
The same argument with fixed sets shows that $f$ is continuous.

The observed and Gaussian-reference statistics can now be written
\[
\begin{split}
 CLR(\beta_m,\Tilde\beta,\hat\Sigma/m,X_m)
 &=f_m(x_m+\hat C_m\Tilde Z_m,\hat\Sigma^{-1},\Tilde Z_m),\\
 CLR(\beta_m,m^{-1/2}\hat{\mathcal Z}_b+\beta_m,\hat\Sigma/m,X_m)
 &=f_m(x_m+\hat C_m\hat{\mathcal Z}_b,
                        \hat\Sigma^{-1},\hat{\mathcal Z}_b).
\end{split}
\]
The second identity holds with $(X_m,\hat\Sigma)$ fixed and
$\hat{\mathcal Z}_b\sim N(0_K,\hat\Sigma_{\beta\beta})$.
Thus it uses exactly the conditional reference distribution in
the subsection.

To study these critical values, let
$\mathcal Z\sim N(0_K,I_K)$ be a standard Gaussian reference vector.
For a positive definite covariance matrix $V$, write
$C(V)=V_{\delta\beta}V_{\beta\beta}^{-1}$ and use the local shorthand
\[
 c_m(x,V)=cv_{1-\alpha}\!\left(
 f_m(x+C(V)V_{\beta\beta}^{1/2}\mathcal Z,
                    V^{-1},V_{\beta\beta}^{1/2}\mathcal Z)\right).
\]
Let $c(x,V)$ be the same critical value with $f$ replacing $f_m$.
The square root is the symmetric positive definite one.
Because $\hat\Sigma_{\beta\beta}^{1/2}\mathcal Z$ has the
conditional law of $\hat{\mathcal Z}_b$, the unadjusted critical
value in Proposition~\ref{P3} is $c_m(x_m,\hat\Sigma)$.

Only a lower bound on limiting critical values is needed.
For any deterministic $(\bar x_m,V_m)\to(x,V)$ with $V$ positive
definite, continuity of matrix inversion and the square root,
together with \eqref{clr:sequential-map}, gives, 
$$f_m(\bar x_m+C(V_m)V_{m,\beta\beta}^{1/2}\mathcal Z,
                         V_m^{-1},V_{m,\beta\beta}^{1/2}\mathcal Z)
 \rightarrow
 f(x+C(V)V_{\beta\beta}^{1/2}\mathcal Z,
                         V^{-1},V_{\beta\beta}^{1/2}\mathcal Z),$$
almost surely (using the same $\mathcal Z$ throughout).

Let $F_{x,V}$ denote the distribution function of the variable on
the right. For any $t<c(x,V)$, choose a continuity point $a$ of
$F_{x,V}$ with $t<a<c(x,V)$. The definition of the lower
$1-\alpha$ quantile gives $F_{x,V}(a)<1-\alpha$.
Weak convergence implies that the finite-$m$ reference distribution
function is also below $1-\alpha$ at $a$ for all sufficiently large
$m$. Thus $c_m(\bar x_m,V_m)\ge a$.
Letting $t$ increase to $c(x,V)$ proves
\begin{equation}\label{clr:quantile-lower-bound}
 \liminf_m c_m(\bar x_m,V_m)\ge c(x,V).
\end{equation}
No density bound or strict crossing of the quantile level is needed.

Apply the Skorokhod representation theorem to the joint convergence
of $(\Tilde Z_m,x_m,\hat\Sigma)$. On the resulting probability
space, these arguments converge almost surely to
$(\mathcal Z_b,\mathcal X,\Sigma)$ with their original laws
preserved. The covariance matrices are eventually positive
definite almost surely. By
\eqref{clr:sequential-map}--\eqref{clr:quantile-lower-bound},
\[
\begin{split}
 f_m(x_m+\hat C_m\Tilde Z_m,\hat\Sigma^{-1},\Tilde Z_m)
 \rightarrow
 f(\mathcal X+C\mathcal Z_b,\Sigma^{-1},\mathcal Z_b),
 \liminf_m c_m(x_m,\hat\Sigma)\ge c(\mathcal X,\Sigma).
\end{split}
\]
Write $T_m$ and $T$ for the statistics on the first line, and
$q_m=c_m(x_m,\hat\Sigma)$ and $q=c(\mathcal X,\Sigma)$ for the
critical values. Extend $T_m,q_m$ by fixed finite values off the
positive-definite event. Its vanishing probability does not affect
limiting acceptance probabilities.
On $\{T\le q\}$, almost surely there is a finite index after which
$T_m\le T+\epsilon/3$ and $q_m\ge q-\epsilon/3$. Hence
\[
 T_m\le T+\epsilon/3\le q+\epsilon/3
       <q-\epsilon/3+\epsilon\le q_m+\epsilon
\]
eventually, including when $T=q$. It follows that
$\mathbf1\{T\le q\}\le
\liminf_m\mathbf1\{T_m\le q_m+\epsilon\}$ almost surely.
Fatou's lemma therefore gives
\[
\begin{split}
 \liminf_m CP_m(\gamma_m)
 &=\liminf_m\mathbb E[\mathbf1\{T_m\le q_m+\epsilon\}]\\
 &\ge\mathbb E[\liminf_m\mathbf1\{T_m\le q_m+\epsilon\}]
 \ge P(T\le q).
\end{split}
\]
Conditional on $\mathcal X$, independence implies that the law of
$T=f(\mathcal X+C\mathcal Z_b,\Sigma^{-1},\mathcal Z_b)$
is precisely the Gaussian reference law whose critical value is
$c(\mathcal X,\Sigma)$. Consequently,
$P(T\le q\mid\mathcal X)\ge1-\alpha$ almost surely.
Integrating gives $P(T\le q)\ge1-\alpha$ and thus
$\liminf_m CP_m(\gamma_m)\ge1-\alpha$, with no restriction on
$P(T=q)$.

In the terminology of Assumption A1 of \cite{ANDREWS2020}, let
$h=(S,S_D,\gamma^*)$ range over the limits attainable by the preceding
simultaneous extraction, and set $CP^-(h)=P(T\le q)$ and
$CP^+(h)=1$. The argument gives $ CP^-(h)\le\liminf_m CP_m(\gamma_m)
 \le\limsup_m CP_m(\gamma_m)\le CP^+(h).$ Thus the coverage probabilities need not converge, but their lower
limits are at least $1-\alpha$. Therefore, Assumption A1 of \cite{ANDREWS2020} holds with $h$ ranging over the attainable
triples $(S,S_D,\gamma^*)$ and with the bounds $CP^-(h)$ and
$CP^+(h)$ defined above. In particular, $CP^-(h)\ge1-\alpha$
for every such $h$. Therefore, by Theorem 2.1(a) of \cite{ANDREWS2020},
$AsySz=\liminf_n\inf_{\gamma\in\Gamma}CP_n(\gamma)\ge1-\alpha$.
\end{proof}
\end{journalappendixmath}

\noindent\textbf{Proof of Proposition~\ref{P5}}
\begin{journalappendixmath}
\begin{proof}
Fix $\{\gamma_n\}_{n\in\mathbb N}\in\Gamma(\gamma^*)$.
By Assumptions~\hyperref[A3]{3.2} and~\hyperref[A3]{3.4}, both
$\hat M_\theta$ and $\partial_\theta\hat M(\theta_n,\psi_n,\eta_n)$
converge in probability to $M_\theta$.
Since $M_\theta$ has full column rank by Assumption~\hyperref[A3]{3.4},
$\hat L=(\hat M_\theta'\hat M_\theta)^{-1}\hat M_\theta'
\overset{p}{\to}L=(M_\theta'M_\theta)^{-1}M_\theta'$.
Using the expansion in Assumption~\hyperref[A3]{3.2} in the one-step
update $\widetilde\theta=\hat\theta-
\hat L\hat M(\hat\theta,\hat\psi,\hat\eta)$ gives
\[
 \sqrt n(\widetilde\theta-\theta_n)
 =(I_J-\hat L\,\partial_\theta\hat M(\theta_n,\psi_n,\eta_n))\sqrt n(\hat\theta-\theta_n)
 -\hat L\sqrt n\,\hat M(\theta_n,\psi_n,\eta_n)
 +o_p(1).
\]
Since $LM_\theta=I_J$ and
$\sqrt n(\hat\theta-\theta_n)=O_p(1)$ by Assumption~\hyperref[A3]{3.1},
the first term is $o_p(1)$.
Assumption~\hyperref[A3]{3.3} gives the $N(0,V_M(\gamma^*))$ limit of the
oracle moment. Consequently,
\[
 \sqrt n(\widetilde\theta-\theta_n)
 =-L\sqrt n\,\hat M(\theta_n,\psi_n,\eta_n)+o_p(1)
 \Rightarrow N(0,LV_M(\gamma^*)L').
\]
Finally, Assumption~\hyperref[A3]{3.5} gives
$\hat V_M(\hat\theta)\overset{p}{\to}V_M(\gamma^*)$. Together with
$\hat L\overset{p}{\to}L$, this yields
$\hat\Sigma=\hat L\hat V_M(\hat\theta)\hat L'
\overset{p}{\to}LV_M(\gamma^*)L'=\Sigma(\gamma^*)$.
\end{proof}
\end{journalappendixmath}

\noindent\textbf{Proof of Proposition~\ref{mt:proposition}}

\begin{journalappendixmath}
\begin{proof}
Fix an arbitrary sequence $\{\gamma_G\}_{G\in\mathbb N}=\{(\theta_G,F_G)\}_{G\in\mathbb N}\in\Gamma(\gamma^*).$ Within the proof, write $V_i=V_{i,n}$ and suppress the experiment index on population expectations. The space $\mathcal K$ is compact metrizable. To establish compactness of $\mathcal C$, take any sequence in that set. Each coordinate lies in $[-B_c,B_c]$, so successive subsequence extraction and a diagonal argument give a subsequence converging coordinatewise to a sequence $b$. Passing to limits in each finite sum gives $\sum_{j=1}^k|b_j|\le B_c$ and $\sum_{j=k+1}^{m}|b_j|\le a_k$ for every $m>k$. Letting $m\to\infty$ shows $b\in\mathcal C$. The head-and-tail inequality above then gives $\ell_1$ convergence of the subsequence. Hence $\mathcal C$ is sequentially compact, and therefore compact in the metric $d_1$. Probability measures on $\mathcal K$ form a compact metrizable space under weak convergence. Each population Gram inequality in Condition~\ref{mt:condition}(i) is closed, because its integrand is bounded and continuous. The assignment inequalities are closed in the coefficient topology, as are the finite-prefix sparsity restrictions. Together with compactness of the error-law index space $\mathcal H$, these facts imply that $\mathcal W$ is closed in a compact product space, and $\Gamma=\Theta\times\mathcal W$ satisfies Assumption~\ref{A1}. This product structure also ensures that no constraint links $\theta$ to the high-dimensional nuisance block.

The functions $e_a$ are uniform limits of continuous finite-coordinate functions, since $\sup_x|e_a(x)-e_{a,n}(x)|\le B_X a_{P_n}$. Thus $M_\theta$ is continuous in $\gamma$, and $M_{\theta,n}(\gamma)\to M_\theta(\gamma)$ uniformly on $\Gamma$. For $h\in\mathbb R^J$, conditional treatment variance satisfies
\[
 h'\operatorname{Var}(D\mid X)h
 =\sum_{0\le a<b\le J}e_{a,n}(X)e_{b,n}(X)(h_a-h_b)^2
 \ge\underline\pi^2\left\lVert h\right\rVert_2^2,
 \qquad h_0=0.
\]
In particular, $M_{\theta,n},M_\theta\succeq\underline\pi^2I_J$. Along $\{\gamma_n\}_{n\in\mathbb N}\in\Gamma(\gamma^*)$, $M_{\theta,n}(\gamma_n)\to M_\theta(\gamma^*)$.

I give the rate argument also for the joint outcome LASSO regression defined in the subsection, since its treatment coefficients are unpenalized and constrained. Write $Z=(D',X')'$. For any $(a,b)$, conditional mean-zero residualization gives
$\mathbb{E}(D'a+X'b)^2 =\mathbb{E}(V'a)^2+\mathbb{E}\{X'(b+\eta_n'a)\}^2 \ge\underline\pi^2\left\lVert a\right\rVert_2^2+c_X\left\lVert b+\eta_n'a\right\rVert_2^2.$
The row $\ell_1$ bounds on $\eta_n$ imply a uniform operator-norm bound, so the last expression is at least $c_Z(\left\lVert a\right\rVert_2^2+\left\lVert b\right\rVert_2^2)$ for some $c_Z>0$. This also makes $(\theta_n,\psi_n)$ the unique population least-squares minimizer. The error-score bound in Condition~\ref{mt:condition}(v) and bounded-regressor concentration imply, for the stated tuning constants, an event with probability tending to one on which
\[
 \left\lVert \frac1n\sum_{i=1}^n Z_i\varepsilon_i\right\rVert_\infty\le\lambda_\psi/2,
 \quad\max_a\left\lVert \frac1n\sum_{i=1}^n X_iV_{ia}\right\rVert_\infty\le\lambda_\eta/2,
\]
$\left\lVert \frac1n\sum_{i=1}^n Z_iZ_i'-\mathbb{E} ZZ'\right\rVert_{\max} =O\bigl(\sqrt{\log(P_n+n)/n}\bigr).$
The first inequality follows from the error-score bound in Condition~\ref{mt:condition}(v). The remaining bounds follow from bounded-variable concentration and union bounds over at most $(P_n+J)^2$ entries, with constants uniform over $\Gamma$. They involve only the treatment and design variables.

Let $\Delta^L=((\hat\theta^{\,L}-\theta_n)',(\hat\psi^{\,L}-\psi_n)')'$, write $\Delta=\Delta^L$, and let $T$ contain all $J$ treatment coordinates, the intercept coordinate, and the support of $\psi_n$. All unpenalized coordinates belong to $T$, and $|T|\le J+1+s_{\psi,n}$. The basic inequality for the joint outcome LASSO regression, comparing with the feasible truth and applying the score bound also to the unpenalized coordinates, yields $ \frac1n\sum_{i=1}^n(Z_i'\Delta)^2
   +\lambda_\psi\left\lVert \Delta_{T^c}\right\rVert_1
 \le3\lambda_\psi\left\lVert \Delta_T\right\rVert_1.$ Thus $\left\lVert \Delta\right\rVert_1\le4\sqrt{|T|}\left\lVert \Delta\right\rVert_2$. On this cone, the sample Gram quadratic form is bounded below by
\[
 \left(c_Z-16|T|\left\lVert \frac1n\sum_{i=1}^n Z_iZ_i'-\mathbb{E} ZZ'\right\rVert_{\max}\right)\left\lVert \Delta\right\rVert_2^2
 \ge(c_Z/2)\left\lVert \Delta\right\rVert_2^2
\]
with probability tending to one, by the sparsity rate restriction in Condition~\ref{mt:condition}(iii). Combining these inequalities gives
\[
 \left\lVert \Delta^L\right\rVert_2=O_p(\sqrt{J+s_{\psi,n}}\lambda_\psi),\qquad
 R_L^2:=\frac1n\sum_{i=1}^n(Z_i'\Delta^L)^2
       =O_p((J+s_{\psi,n})\lambda_\psi^2).
\]
For the refit, put $\Delta^{PL}=((\hat\theta^{\,PL}-\theta_n)',(\hat\psi-\psi_n)')'$ and $R_{PL}^2=\frac1n\sum_{i=1}^n(Z_i'\Delta^{PL})^2$.
Its support is contained in the treatment coordinates together with $\hat S\cup\operatorname{supp}(\psi_n)$, of size $d_n=O_p(J+s_{\psi,n})$. Entrywise Gram concentration and the sparsity rate restriction in Condition~\ref{mt:condition}(iii) imply, uniformly over vectors on such supports,
$R_{PL}^2\ge(c_Z/2)\left\lVert \Delta^{PL}\right\rVert_2^2$
with probability tending to one. Indeed, the Gram perturbation is bounded by $d_n O_p(\sqrt{\log(P_n+n)/n})=o_p(1).$ The selected model contains the LASSO fit, although it need not contain the truth. Thus the refit criterion is no larger than the LASSO fit's unpenalized criterion, and expansion around the truth yields
\[
 R_{PL}^2
 \le R_L^2+2\left(\frac1n\sum_{i=1}^nZ_i\varepsilon_i\right)'(\Delta^{PL}-\Delta^L)
 \le R_L^2+\lambda_\psi(\left\lVert \Delta^{PL}\right\rVert_1+\left\lVert \Delta^L\right\rVert_1).
\]
Using $\left\lVert \Delta^{PL}\right\rVert_1\le\sqrt{d_n}\left\lVert \Delta^{PL}\right\rVert_2$, the preceding eigenvalue bound, and the LASSO bounds gives
$R_{PL}=O_p(\sqrt{J+s_{\psi,n}}\lambda_\psi)$ and the same rate for $\left\lVert \Delta^{PL}\right\rVert_2$. Consequently,
\begin{equation}\label{mt:oracle}
 \left\lVert \hat\psi-\psi_n\right\rVert_1=O_p((J+s_{\psi,n})\lambda_\psi),\qquad
 \frac1n\sum_{i=1}^n\bigl(X_i'(\hat\psi-\psi_n)\bigr)^2
       =O_p((J+s_{\psi,n})\lambda_\psi^2).
\end{equation}
For the second conclusion, use the prediction bound for $Z'\Delta^{PL}$, the Euclidean error bound just established, and bounded $D$. The same basic-inequality argument for each regression \eqref{mt:node}, including its unpenalized intercept in the active set and using the population Gram bound in Condition~\ref{mt:condition}(i), gives
\begin{equation}\label{mt:etarate}
 \left\lVert \operatorname{vec}(\hat\eta-\eta_n)\right\rVert_1
       =O_p((J+s_{\eta,n})\lambda_\eta).
\end{equation}
The rate condition implies both $\ell_1$ errors are $o_p(1)$. I use these rates below to bound the effect of estimating the nuisance parameters.

Since $\mathbb{E}[D\mid X]=\eta_nX$,
$D_{\theta\psi}Q_n-\eta_nD_{\psi\psi}Q_n =\mathbb{E}[DX']-\eta_n\mathbb{E}[XX']=0.$
Set $\Delta_\psi=\hat\psi-\psi_n$ and $\Delta_\eta=\hat\eta-\eta_n$. The exact identity $\hat M(\theta_n,\hat\psi,\hat\eta)
       -\hat M(\theta_n,\psi_n,\eta_n) =\Delta_\eta\left(\frac1n\sum_{i=1}^n X_i\varepsilon_i\right)
       +\left(\frac1n\sum_{i=1}^n(D_i-\hat\eta X_i)X_i'\right)\Delta_\psi$ and the KKT conditions (zero sample residual mean for the intercept and an absolute correlation bound $\lambda_\eta$ for every other control) give the inequalities $\left\lVert \frac1n\sum_{i=1}^n X_i(D_{ia}-X_i'\hat\eta_a)\right\rVert_\infty\le\lambda_\eta$. Let $R_n=\hat M(\theta_n,\hat\psi,\hat\eta)
-\hat M(\theta_n,\psi_n,\eta_n)$. Applying these correlation bounds
and the error-score bound to the two terms above, and using
\eqref{mt:oracle} and \eqref{mt:etarate}, gives $ \|R_n\|_2
 =O_p\bigl((J+s_{\eta,n})\lambda_\eta\sqrt{\log(P_n+n)/n}
           +(J+s_{\psi,n})\lambda_\eta\lambda_\psi\bigr)
 =o_p(n^{-1/2})$ by Condition~\ref{mt:condition}(iii). This calculation uses the coefficient bound for the outcome refit and the KKT inequalities for the treatment regressions.

At the truth, $ \hat M=-\frac1n\sum_{i=1}^n V_i\varepsilon_i$ and $\partial_\theta\hat M=\frac1n\sum_{i=1}^n V_iD_i'.$ The score CLT in Condition~\ref{mt:condition}(v), boundedness of $V_i,D_i$, and continuity of $M_\theta$ give $\sqrt n\hat M(\theta_n,\psi_n,\eta_n)
   \Rightarrow N(0,V_M(\gamma^*))$ and $\partial_\theta\hat M(\theta_n,\psi_n,\eta_n)
   \rightarrow_p M_\theta(\gamma^*).$ These verify Assumptions~\hyperref[A3]{3.3}--\hyperref[A3]{3.4}.

Write $G_n=\partial_\theta\hat M(\theta_n,\psi_n,\eta_n)
=n^{-1}\sum_{i=1}^n V_iD_i'$ for the sample Jacobian at the true nuisance
parameters. The feasible moment is affine in $\theta$, with Jacobian
$\hat M_\theta=n^{-1}\sum_{i=1}^n(D_i-\hat\eta X_i)D_i'$.
Since $X_i$ and $D_i$ are bounded and $J$ is fixed,
\[
 \|\hat M_\theta-G_n\|_2
 =\left\|\Delta_\eta\,\frac1n\sum_{i=1}^nX_iD_i'\right\|_2
 \le C\|\operatorname{vec}(\Delta_\eta)\|_1=o_p(1)
\]
by \eqref{mt:etarate}. This is the Jacobian approximation in
Assumption~\hyperref[A3]{3.2}. In particular,
$\hat M_\theta\to_p M_\theta(\gamma^*)$.

Let $m_n=\hat M(\theta_n,\hat\psi,\hat\eta)$, which is $O_p(n^{-1/2})$ by the bound on $R_n$ and the oracle CLT. Let $r_n=o_p(n^{-1})$ denote the nonnegative optimization error in \eqref{mt:gmm}. The score is affine in $\theta$, so feasibility of $\theta_n$ and \eqref{mt:gmm} give
\[
 \left\lVert \hat M_\theta(\hat\theta-\theta_n)\right\rVert_2
 \le\left\lVert \hat M(\hat\theta,\hat\psi,\hat\eta)\right\rVert_2+\left\lVert m_n\right\rVert_2
 \le2\left\lVert m_n\right\rVert_2+\sqrt{r_n}=O_p(n^{-1/2}).
\]
The smallest singular value of $\hat M_\theta$ is bounded away from zero with probability tending to one. Thus $\sqrt n(\hat\theta-\theta_n)=O_p(1)$, verifying Assumption~\hyperref[A3]{3.1} even on the boundary. For the moment expansion in Assumption~\hyperref[A3]{3.2}, affine
dependence on $\theta$ gives the exact identity
\begin{align*}
 \hat M(\hat\theta,\hat\psi,\hat\eta)
  -\hat M(\theta_n,\psi_n,\eta_n)-G_n(\hat\theta-\theta_n)
 =R_n+(\hat M_\theta-G_n)(\hat\theta-\theta_n).
\end{align*}
The first term is $o_p(n^{-1/2})$. The second is
$o_p(1)O_p(n^{-1/2})=o_p(n^{-1/2})$ by the Jacobian bound and the
rate just proved. This verifies the remaining part of
Assumption~\hyperref[A3]{3.2}. Affine moments also give the exact cancellation
$\widetilde\theta =\hat M_\theta^{-1}\left(\frac1n\sum_{i=1}^n(D_i-\hat\eta X_i)(Y_i-X_i'\hat\psi)\right),$
whenever $\hat M_\theta$ is invertible.

The true $V$ are bounded, and $\sup_i\left\lVert \hat v_i-V_i\right\rVert=o_p(1)$ follows from \eqref{mt:etarate} and bounded controls. Hence $\sup_i\left\lVert \hat v_i\right\rVert=O_p(1)$. The bounds in \eqref{mt:oracle} and the root-$n$ rate just proved give $\frac1n\sum_{i=1}^n(\hat u_i-\varepsilon_i)^2=o_p(1)$. The uniform second-moment bound implies $\frac1n\sum_{i=1}^n\varepsilon_i^2=O_p(1)$ by Markov's inequality. It follows that
$\frac1n\sum_{i=1}^n\left\lVert \hat v_i\hat u_i-V_i\varepsilon_i\right\rVert_2^2=o_p(1).$
Cauchy--Schwarz and the covariance law of large numbers in Condition~\ref{mt:condition}(v) imply
$\hat V_M(\hat\theta)=\hat V_M\to_p V_M(\gamma^*)$,
verifying Assumption~\hyperref[A3]{3.5}. All five parts of
Assumption~\ref{A3} and the population orthogonality equation now hold
along the arbitrary drifting sequence. Proposition~\ref{P5} therefore gives
$\sqrt n(\widetilde\theta-\theta_n)\Rightarrow N(0,\Sigma(\gamma^*))$
and $\hat\Sigma\to_p\Sigma(\gamma^*)$.
\end{proof}
\end{journalappendixmath}

\textbf{Lemma A1}\label{LA1} Let $d_1$ and $d_2$ be nonnegative integers, and let $d = d_1 + d_2$. Let $\{z_m\}_{m=1}^\infty$ be a sequence in $\mathbb{R}^{d_1}$ such that $z_m \to z$, and let $\{x_m\}_{m=1}^\infty$ be a sequence in $\mathbb{R}^{d_2}$ such that $x_m \to x$. Let $\{Y_m\}_{m=1}^\infty$ be a sequence of positive definite matrices in $\mathbb{R}^{d \times d}$ such that $Y_m \to Y$, where $Y$ is also positive definite. Let $\{S_m\}_{m=1}^\infty$ be a sequence of subsets of $\mathbb{R}^{d_2}$ that converges to a non-empty set $S \subset \mathbb{R}^{d_2}$ in the Painlev\'{e}-Kuratowski sense. Define the quadratic forms $h_m, h : \mathbb{R}^{d_2} \to \mathbb{R}$ by
\begin{align*}
    h_m(b) = \begin{pmatrix} z_m \\ x_m - b \end{pmatrix}^\top Y_m \begin{pmatrix} z_m \\ x_m - b \end{pmatrix},
    h(b) = \begin{pmatrix} z \\ x - b \end{pmatrix}^\top Y \begin{pmatrix} z \\ x - b \end{pmatrix}.
\end{align*}
Then, the sequence of infima converges: $\lim_{m \to \infty} \inf_{b \in S_m} h_m(b) = \inf_{b \in S} h(b).$

\begin{journalappendixmath}
    \begin{proof}
        Let $v_m = \inf_{b \in S_m} h_m(b)$ and $v = \inf_{b \in S} h(b)$. Since $S \neq \emptyset$ and $Y$ is positive definite, we have $0 \leq v < \infty$. I proceed by showing that $\limsup_{m \to \infty} v_m \leq v$ and $\liminf_{m \to \infty} v_m \geq v$. First, I establish the upper bound. Let $b \in S$ be arbitrary. By the inner limit property of Painlev\'{e}-Kuratowski convergence, there exists a sequence $\{b_m\}_{m=1}^\infty$ such that $b_m \in S_m$ for all $m$, and $b_m \to b$. By the definition of the infimum, 
\begin{equation}\label{eq:upper_bound_ineq}
    v_m \leq h_m(b_m) = \begin{pmatrix} z_m \\ x_m - b_m \end{pmatrix}^\top Y_m \begin{pmatrix} z_m \\ x_m - b_m \end{pmatrix}.
\end{equation}
Since $z_m \to z$, $x_m \to x$, $b_m \to b$, and $Y_m \to Y$, the continuity of vector and matrix operations implies that $h_m(b_m) \to h(b)$. Taking the limit superior of \eqref{eq:upper_bound_ineq} yields
$\limsup_{m \to \infty} v_m \leq \lim_{m \to \infty} h_m(b_m) = h(b).$
Because this holds for every $b \in S$, taking the infimum over all $b \in S$ provides the upper bound: $\limsup_{m \to \infty} v_m \leq v$. Next, I establish the lower bound. Extract a subsequence $\{v_{m_k}\}_{k=1}^\infty$ such that $\lim_{k \to \infty} v_{m_k} = \liminf_{m \to \infty} v_m$. For each $k$, choose a point $b_{m_k} \in S_{m_k}$ such that
\begin{equation}\label{eq:lower_bound_seq}
    h_{m_k}(b_{m_k}) \leq v_{m_k} + \frac{1}{k}.
\end{equation}
From our upper bound result, the sequence $\{v_{m_k}\}$ is bounded above, which implies that the sequence $\{h_{m_k}(b_{m_k})\}$ is also bounded. Since $Y$ is positive definite over the full space $\mathbb{R}^d$, its minimum eigenvalue $\lambda_{\min}(Y)$ is strictly positive. The convergence $Y_m \to Y$ ensures that the matrices $Y_{m_k}$ are uniformly coercive for sufficiently large $k$; that is, there exists a constant $c > 0$ such that 
\begin{equation}\label{eq:coercivity}
    h_{m_k}(b_{m_k}) \geq c \left\| \begin{pmatrix} z_{m_k} \\ x_{m_k} - b_{m_k} \end{pmatrix} \right\|^2 = c \left( \|z_{m_k}\|^2 + \|x_{m_k} - b_{m_k}\|^2 \right) \geq c \|x_{m_k} - b_{m_k}\|^2.
\end{equation}
Given that $\{h_{m_k}(b_{m_k})\}$ is bounded and $\{x_{m_k}\}$ is bounded (since it converges to $x$), the inequality in \eqref{eq:coercivity} dictates that the sequence $\{b_{m_k}\}$ must also be bounded. By the Bolzano-Weierstrass theorem, $\{b_{m_k}\}$ possesses a convergent sub-subsequence. Without loss of generality, assume the sequence $\{b_{m_k}\}$ itself converges to some limit $\hat{b}$. By the outer limit property of Painlev\'{e}-Kuratowski convergence, the conditions $b_{m_k} \in S_{m_k}$ and $b_{m_k} \to \hat{b}$ guarantee that $\hat{b} \in S$. Taking the limit as $k \to \infty$ of the functions evaluated at this subsequence yields $\lim_{k \to \infty} h_{m_k}(b_{m_k}) = h(\hat{b})$. Because $\hat{b} \in S$, it follows that $h(\hat{b}) \geq \inf_{b \in S} h(b) = v$. Applying this to \eqref{eq:lower_bound_seq}, we have
\begin{equation*}
    \liminf_{m \to \infty} v_m = \lim_{k \to \infty} v_{m_k} \geq \lim_{k \to \infty} \left( h_{m_k}(b_{m_k}) - \frac{1}{k} \right) = h(\hat{b}) \geq v.
\end{equation*}
Combining the upper and lower bounds yields $v \leq \liminf_{m \to \infty} v_m \leq \limsup_{m \to \infty} v_m \leq v$, which implies $\lim_{m \to \infty} v_m = v$ and concludes the proof.
    \end{proof}
\end{journalappendixmath}

\end{document}